\documentclass[lettersize,journa]{IEEEtran}
\usepackage{amsthm}
\usepackage{amsmath}
\usepackage{amsfonts}
\usepackage{amssymb}
\usepackage{mathrsfs}
\usepackage{bm}
\usepackage{graphicx}
\usepackage{subfigure}
\usepackage{stfloats}
\usepackage{booktabs}
\usepackage{hyperref}
\usepackage{cite}
\usepackage{cases}
\usepackage{tabularx}
\usepackage{textcomp}
\usepackage[ruled,linesnumbered,vlined]{algorithm2e}

\begin{document}
\title{Energy-Efficient Integrated Communication and Computation via Non-Terrestrial Networks\\with Uncertainty Awareness}

\author{Xiao Tang, Yudan Jiang, Ruonan Zhang, Qinghe Du, Jinxin Liu, and Naijin Liu% <-this % stops a space
\thanks{X. Tang is with School of Information and Communication Engineering, Xi’an Jiaotong University, Xi’an 710049, China, and also with Research \& Development Institute of Northwestern Polytechnical University in Shenzhen, Shenzhen 518063, China. (e-mail: tangxiao@xjtu.edu.cn)}% <-this % stops a space
\thanks{Y. Jiang and R. Zhang are with School of Electronics and Information, Northwestern Polytechinical University, Xi'an 710072, China.}% <-this % stops a space
\thanks{Q. Du is with School of Information and Communication Engineering, Xi'an Jiaotong University, Xi'an 710049, China.}
\thanks{J. Liu is with School of Mechanical Engineering, Xi’an Jiaotong University, Xi’an 710049, China.}% <-this % stops a space
\thanks{N. Liu is with Qian Xuesen Laboratory of Space Technology, China Academy of Space Technology, Beijing 100094, China.}% <-this % stops a space
}
\maketitle

\begin{abstract}
Non-terrestrial network (NTN)-based integrated communication and computation empowers various emerging applications with global coverage. Yet this vision is severely challenged by the energy issue given the limited energy supply of NTN nodes and the energy-consuming nature of communication and computation. In this paper, we investigate the energy-efficient integrated communication and computation for the ground node data through a NTN, incorporating an unmanned aerial vehicle (UAV) and a satellite. We jointly consider ground data offloading to the UAV, edge processing on the UAV, and the forwarding of results from UAV to satellite, where we particularly address the uncertainties of the UAV-satellite links due to the large distance and high dynamics therein. Accordingly, we propose to minimize the weighted energy consumption due to data offloading, UAV computation, UAV transmission, and UAV propulsion, in the presence of angular uncertainties under Gaussian distribution within the UAV-satellite channels. The formulated problem with probabilistic constraints due to uncertainties is converted into a deterministic form by exploiting the Bernstein-type inequality, which is then solved using a block coordinate descent framework with algorithm design. Simulation results are provided to demonstrate the performance superiority of our proposal in terms of energy sustainability, along with the robustness against uncertain non-terrestrial environments.
\end{abstract}

\begin{IEEEkeywords}
Integrated communication and computation, non-terrestrial network, unmanned aerial vehicle, energy, uncertainty
\end{IEEEkeywords}

\section{Introduction}

The future 6G networks are expected to provide global coverage beyond the earth surface to non-terrestrial space, and thus bringing forth a paradigm shift toward a seamlessly connected world~\cite{6G-1}. Towards this vision, the non-terrestrial network (NTN) incorporating various aerial platforms and satellites plays a pivotal role in enabling and enhancing the globally ubiquitous access~\cite{NTN-2}. Given the significant advantages in terms of flexibility and cost efficiency, the NTN extends the coverage to the areas which are economically unfriendly and challenging with conventional network infrastructure. Thus, the enhanced connectivity effectively supports massive access to mobile devices with emerging services and applications~\cite{NTN-cov}. Along with the enhanced coverage of the network, the capabilities of the nodes in NTNs are also witnessed with growing strength, and thus enables integrated communication and computation to realize the full potential of NTNs~\cite{CC-1,aa}. In this regard, the network can not only transmit data but also process the data at the network edge, and thus enable a plethora of new applications and functionalities. With data processed near the sources, it reduces the latency due to the transmission to central servers, and also saves the bandwidth and resources to improve the network operating efficiency. Moreover, given the flexible deployment of NTNs, the integration of communication and computation enables dynamic data processing, allowing the network to scale efficiently to satisfy the varying user requirements~\cite{CC-3}.

Towards NTN-supported 6G vision, one of the most critical issues is the energy~\cite{NTN-2}. Basically, the nodes in NTN, such as unmanned aerial vehicles (UAVs), high-altitude platforms, and satellites, rely on onboard energy sources and thus are of strict energy constraints. Meanwhile, the users of NTN particularly include the remote ones beyond cellular coverage, which are usually self-sustained and energy-sensitive~\cite{E-1,E-2}. In contrast, the communication and computation demands in NTN can be rather energy-intensive to facilitate various emerging applications, which situation becomes even more challenging given the complicated architecture and dynamics of NTN and the massive nodes accommodated~\cite{E-3}. In this respect, the existing work mostly addresses the energy issue from the perspective of UAV edge and especially concentrates on computation~\cite{E-4,E-5}. Despite the critical role of UAVs, it may not be able to provide a comprehensive framework to address communication and computation in the NTN context. Rather, the energy issue needs to be addressed by jointly considering different characteristics of the roles in NTN as well as the communication and computation requirements towards effective solutions. In this respect, the energy-efficient designs improve the sustainability of both individual nodes as well as the whole network, and also contribute to the cost-saving network operations and green economy.

Meanwhile, the efficient operations of NTN are significantly challenged by the uncertainties in the network~\cite{U-1}. In the context of NTN, the main contributors of uncertainties may include the UAV trajectory dynamics, the satellite link variability, user mobility, resources fluctuations, and channel estimation errors. These uncertainties affect the performance of NTN in different scales and aspects, for which the influence of uncertainties needs to be tackled effectively~\cite{U-2}. In this respect, the existing work mainly considers the channel uncertainties due to the non-line-of-sight small-scale fading, where the worst-case countermeasure is addressed for robust designs~\cite{U-3,U-4}. However, due to the high dynamics of NTN, the small-scale fading may not always account for the principal contributor of uncertainties. Also, the error bound information used to tackle the worse case may not always be accurately obtained given the complicated uncertain environment. Consequently, it is essential to extract the potentially most significant factor of uncertainties for certain environment, with accurate model of the associated errors so as to tackle it effectively.

In this paper, we propose to exploit NTN-based integrated communication and computation to facilitate ground services, where we intend for energy-efficient design to save the energy for the space, air, and ground roles in the network and achieve robustness in the presence of uncertainties. Specifically, the main contribution can be summarized as follows:
\begin{itemize}
	\item We propose an integrated communication and computation framework under NTN environment, where a UAV flies approaching the ground nodes to process the offloaded data, with computation results forwarded to a satellite for remote uses. In this regard, we jointly investigate the ground node offloading, UAV computation and transmission, and the UAV trajectory for energy-efficient strategy design.
	% \item We formulate the problem to minimize the weighted-sum energy due to ground data offloading to the UAV, UAV computation and results forwarding to the satellite, as well as the UAV propulsion. Moreover, we particularly address the uncertain angular information of the UAV-satellite links with Gaussian distribution and establish the chance-constrained robust performance guarantee.
	\item We particularly address the uncertainties associated with the UAV-satellite links, where the angular information is modeled with a Gaussian distribution. Then, we formulate the problem to minimize the weighted-sum energy due to ground data offloading, UAV computation and results forwarding to the satellite, and the UAV propulsion, with uncertainty-induced chance constraints to ensure robust performance.
	\item We reformulate the optimization with probabilistic-form constraints by employing the Bernstein-type inequality to achieve its deterministic counterpart, which is decomposed into the subproblems to tackle the offloading and scheduling, slot allocation, transmission, and trajectory, respectively. The subproblems are then solved and iterated within a block coordinate descent framework to obtain the energy-efficient integrated communication and computation solution with robustness provisioning.
\end{itemize}

The rest of this paper is organized as follows. In Sec.~\ref{sec2}, we review the related work. We in Sec.~\ref{sec3} and Sec.~\ref{sec4} introduce the NTN-based system model and present the energy minimization problem for integrated communication and computation with uncertainties, respectively. Sec.~\ref{sec5} analyzes the formulated problem with proposed algorithm design and Sec.~\ref{sec5} provides the simulation results. Finally, this paper is concluded in Sec.~\ref{sec6}.

\section{Related Work} \label{sec2}

As one of the essentials for future 6G wireless networks, NTN with unique capabilities such as global coverage, high mobility, and strong resilience make it an ideal complement for conventional terrestrial networks~\cite{NTN-2}. As such, NTN has attracted the research interest in wide aspects, including the network deployment~\cite{NTN-deploy}, function placement~\cite{NTN-place}, constellation design~\cite{NTN-con}, communication optimization~\cite{NTN-com}, security enhancement~\cite{ee}, and trust management~\cite{cc}. Built upon such enhanced network infrastructure, the integrated communication and computation empowers the network with the potential of prosperous applications~\cite{ICC-intro}. In~\cite{ICC-sel}, the authors consider the satellite-terrestrial edge computing network and investigate the server placement and service deployment issue to minimize the latency and energy consumption. In~\cite{ICC-cf}, the authors propose a HAP-enabled aerial cell-free network to realize edge computing serving massive Internet-of-Things nodes with grant-free access. In~\cite{ICC-drl}, the authors develop a multi-timescale learning scheme to optimize radio resources in NTNs, enabling collaborative decision-making between satellites and user equipment. In~\cite{dd}, the authors investigate the UAV-assisted satellite computing with a game theoretical framework to model the resource competing among ground users. The Nash equilibrium is achieved so as to minimize the overall cost while meeting resource and satellite coverage constraint. In~\cite{bb}, the authors propose a digital twin-assisted UAV deployment strategy along with a hybrid task offloading scheme, facilitating high-fidelity real-world state information for model training.
% In~\cite{ICC-fl}, the authors propose an asynchronous federated learning framework, enhancing computation offloading to improve model training accuracy under NTN environment.
% handover design~\cite{NTN-ho}

Energy issue presents a critical challenge for NTN operations, where the high energy consumption to support various applications necessitates novel solutions to achieve energy-efficient designs~\cite{NTN-cov,cja}. In this context, the popularity of UAVs has made it an attractive assisting role for energy-efficient computations. In this regard, the existing works propose to exploit the dynamic flying of the UAVs as flexible edge servers to provide computation services, where the computation and flying are jointly investigated to improve the energy efficiency, along with the consideration on completion time~\cite{Engy-uav-time}, data queuing~\cite{Engy-uav-traj}, and so forth. Moreover, the UAV-aided computation are also jointly investigated with emerging technologies, such as reconfigurable reflection~\cite{Engy-uav-ris} and energy harvesting~\cite{Engy-uav-eh} to further improve the energy performance. Meanwhile, the satellite counts on single-source energy to support wide-range and long-distance communication and computation, and thus particularly demands energy savings. As such, the satellite computing is jointly investigated with offloading transmission~\cite{Engy-sat-terr}, completion time planning~\cite{Engy-sat-peer}, and task allocation~\cite{Engy-sat-obt} to achieve energy-efficient designs. To further promote NTN applications, it is indispensable to comprehensively consider the features regarding the space, air, and ground roles involved, with jointly design of communication and computation strategies to optimize the system energy performance.

Since the uncertainties originate from multiple sources and widely exist in the NTN environment, proper treatment of uncertainties has also led to devoted research efforts~\cite{U-1,Unct-base}. In this respect, the communication and computation under NTN are investigated incorporating the uncertainties in terms of channel state information~\cite{U-1}, available resources~\cite{Unct-res}, task processing complexity~\cite{Unct-tsk}, to obtain robust approaches. Furthermore, the uncertainties affect the NTN operations and thus induce additional resource consumption, compelling an urgent need for robustness in energy efficient operations. In~\cite{Unct-engy-sec}, the authors propose an energy minimization scheme for secure UAV edge networks under channel uncertainties, achieving robust optimization over various system parameters. In~\cite{Unct-engy-iiot}, the authors address the UAV jittering issue and achieve energy-efficient edge computation. In~\cite{E-2}, the authors consider the environment uncertainty for energy harvesting and introduce a battery-aware energy optimization algorithm for satellite edge computing. In~\cite{U-2}, the authors investigate the energy-constrained computation offloading in a space-air-ground network, addressing task arrival uncertainties towards distributionally robust design. Given the complicated nature of NTN and the uncertainties therein, we need to analyze the particular application scenario to combat the principal uncertain factor to achieve robust energy-efficient communication and computation.

\section{System Model} \label{sec3}

\begin{figure}[t]
  \centering
  \includegraphics[width=6.0cm]{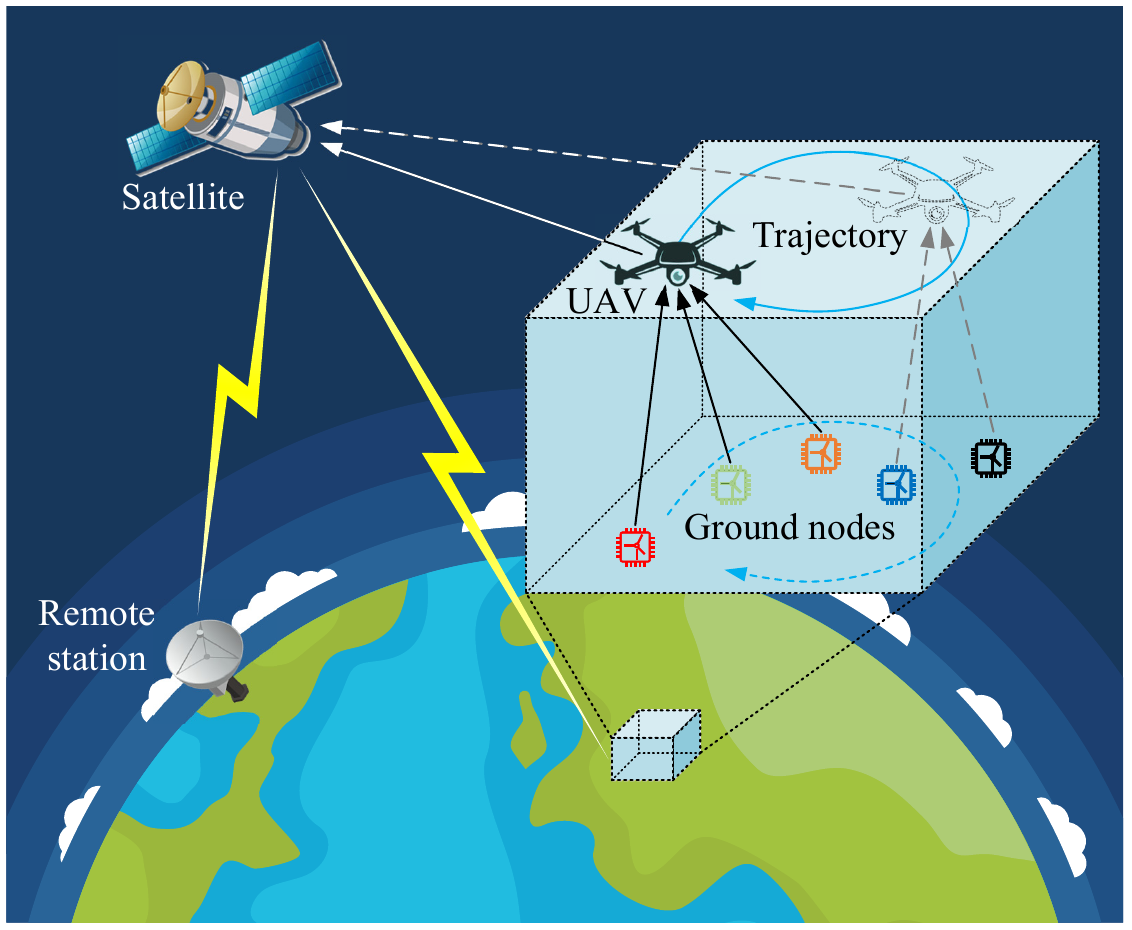}
  \caption{System model.}
  \label{fig:sys}
\end{figure}

We consider an area co-existing $K$ ground nodes, denoted by ${\mathcal{K}} = \left\{ {{ {1,2,3,}} \cdots ,K} \right\}$, where the $k$-th node has the coordinates ${\bm{w}_k} = \left[ {{w}_k^{\left( x \right)},{w}_k^{\left( y \right)},0} \right]$. The nodes are deployed to collect local data in their vicinity to accomplish certain missions. However, given the limited energy and computation capacity of these nodes, data from the ground nodes are offloaded to the NTN for further processing. Particularly, a UAV is dispatched to approach the nodes in this area such that the collected data are offloaded and processed with UAV onboard computation. Additionally, the computation results at the UAV are forwarded to a satellite to facilitate the use at remote stations. The system is illustrated in Fig.~\ref{fig:sys}. As such, the NTN-enabled integrated communication and computation incorporates the data offloading, data computation, and result forwarding as a three-stage process, which effectively tracks the typical application scenarios such as remote monitoring, pipeline inspection, wildlife conservation, etc. Specifically, we expatiate this process with a time-slotted model, where the time horizon $T$ is equally divided into $N$ slots, denoted by ${\mathcal{N}} = \left\{ {1,2,\cdots,N} \right\}$, with each slot having a time length of $T/N$. In this regard, there are $ N+1 $ time instances, denoted by $\bar{\mathcal{N}} = {\mathcal{N}}\cup \{0\} $, for which we use the UAV location at a certain instance as the approximation for the flying over the corresponding slot. We assume the UAV flies at a fixed altitude $H$, and denote the UAV location at time instance-$n$ as ${\bm{q}_n} = \left[ {{q}_n^{\left( x \right)},{q}_n^{\left( y \right)},H} \right]$, where the starting point and finishing point of flying are specified as
\begin{equation} \label{eq:q_sf}
	{\bm{q}_0} = {\bm{q}_S}, \quad {\bm{q}_{N + 1}} = {\bm{q}_F}.
\end{equation}
Meanwhile, the UAV flying is subject to the speed limit as
\begin{equation} \label{eq:speed}
\left\| {{\bm{q}_n} - {\bm{q}_{n - 1}}} \right\| \le {V^{\max }}\frac{T}{N},
\end{equation}
where ${V^{\max }}$ is the highest speed. Meanwhile, given the flying waypoint of the UAV in time slot-$n$, the distance from the UAV to the satellite is denoted by $ d^{\text{(SU)}}_n $. Given the fact that the distance from the satellite to the UAV is significantly larger than that between the UAV and the ground nodes, the UAV-satellite distance is not addressed as an explicit function of the UAV waypoints.

% \subsection{Channel Model}

For the ground-air transmissions corresponding to the data offloading from the ground nodes to the UAV, the channel from node-$k\in\mathcal{K}$ to the UAV in time slot-$n\in\mathcal{N}$ is given as
\begin{equation} \label{eq:h_nk} % equ1 \ref{equ1} 
{h_{n,k}} = \frac{{{\beta _0}}}{{{{\left\| {{\bm{q}_n} - \bm{w}{}_k} \right\|}^2}}}, \quad \forall k \in \mathcal{K},\quad\forall n \in \mathcal{N},
\end{equation}
where ${\beta _0}$ is the signal attenuation at the reference distance, and we assume there is one single antenna at both the ground nodes and the UAV. We use the channel gain at the trajectory waypoints at each time instance to approximate that during the flying in the corresponding time interval.

For the air-space transmissions from UAV to satellite corresponding to the results forwarding process after the data computation at the UAV, where we assume there $M$ receive antennas at the satellite. The UAV-satellite channel propagation incorporates the free-space loss, rain attenuation, antenna gain, and phase effect, as detailed in the followings.

\textit{1) Free-space path loss:}
Given the high altitude of satellite and the long distance of air-space transmissions from UAV to satellite, the free-space signal path loss is given as 
\begin{equation}
\label{equ7} % equ1 \ref{equ1} 
{G_{n}^{\text{(PL)}}} = {\left( {\frac{c }{{4\pi f^{\text{(carr)}} {d^{\left(\text{SU} \right)}_n} }}} \right)^2},
\end{equation}
in time slot-$n$, where $ c $ is the speed of light and $ f^{\text{(carr)}} $ is the carrier frequency. Here we use the UAV-satellite distance and neglect the distance from the beam center to the nodes on the ground as the former dominates the latter.

\textit{2) Rain attenuation:}
As the UAV can be covered by a single satellite beam, the rain attenuation as a slow fading process can be modeled as
\begin{equation} \label{eq:rain} % equ1 \ref{equ1} 
\bm{r}_n = {\varsigma ^2}{\bm{1}_M},  \quad\forall n \in \mathcal{N},
\end{equation}
in the $n$-th time slot, where the power gain $\varsigma^2$ in decibel follows a lognormal distribution.

\textit{3) Antenna gain:}
Generally, the transmit antenna gain, denoted by $ G_{t} $, depends on the elevation angle from the UAV to the satellite, which can be assumed as constant given the relatively small motion of the UAV compared with the large distance of the air-space transmissions, given by $ G_{t}^{\max} $. Meanwhile, the receive antenna gain at the satellite can be obtained as
\begin{equation} \label{eq:G_r} % equ1 \ref{equ1} 
G_r(\phi)=G_{r}^{\max}\left(\frac{J_1(u)}{2u}+36\frac{J_3(u)}{u^3}\right)^2,
\end{equation}
where $G_{r}^{\max}$ is the maximum gain, $ J(\cdot) $ is the Bessel function, $ u = 2.07123\frac{\sin\phi}{\sin\phi_{3\text{dB}}} $, $ \phi $ is the angle between the satellite and antenna boresight, and $ \phi_{3\text{dB}} $ is the corresponding 3-dB angle.

\textit{4) Phase effect:}
Given the line-of-sight transmission from UAV to satellite, the phase of incident signal at the satellite depends on the transmission distance and the antenna separations, which is denoted by $ \bm{\theta}_n = \left[ \theta_{n,m} \right]_{m=1,2,\cdots,M} $ at time slot-$n$.

Based on the discussions on the components above, the UAV-satellite link gain is obtained as
\begin{equation} \label{eq:g_n} % equ1 \ref{equ1} 
{{{\bm{g}}_n}} = \sqrt {\frac{{{G_{n}^{\text{(PL)}}}{G_{t}}{G_{r}}}}{{\kappa B_2t}}} \bm{r}_n^{-\frac{1}{2}} \odot {e^{j{\bm{\theta}_n }}}, \quad \forall n \in \mathcal{N},
\end{equation}
where the noise term $ \kappa B_2t $ is involved as a normalizing factor, with $ \kappa $, $ B_2 $, and $ t $ being the Boltzmann constant, the air-space link bandwidth, and the noise temperature, respectively~\cite{Unct-base}.

\section{Problem Formulation} \label{sec4}

\subsection{Integrated Communication and Computation}
For the proposed NTN-enabled integrated communication and computation for the ground nodes, it is an offloading-computation-forwarding process as elaborated below. Particularly, consider time slot-$n$ as the UAV approximated at $ \bm{q}_n $, the time length $T/N$ is divided into two parts with a coefficient $ \rho_n\in[0,1] $, where $ \rho_n T/N $ is used for the data offloading from the ground nodes to the UAV, and $ (1-\rho_n) T/N $ for the results forwarding from the UAV to the satellite. Here we assume the data computation is conducted while receiving at the UAV, e.g., feeding the data into a pre-trained neural network to produce the results, the computation is conducted in a rather prompt manner and thus it is safe to continue without explicit consideration of computation time. Furthermore, for the data offloading from the ground nodes to the UAV, the ground nodes are scheduled that there is only one active node at at a time. Thus, we use the indicator $ \left\{\chi_{n,k}\right\}_{n\in\mathcal{N},k\in\mathcal{K}} $ with
\begin{equation} \label{eq:chi_con} % equ1 \ref{equ1} 
\left\{ \begin{aligned}
&\chi _{n,k} \in \left\{ {0,1} \right\}, &&\quad \forall k \in \mathcal{K},\:\forall n \in \mathcal{N},\\
&\sum\limits_{k=1}^K {{\chi _{n,k}}} = 1, &&\quad \forall n \in \mathcal{N},
\end{aligned} \right.
\end{equation}
specifying the scheduling, where $ \chi_{n,k} =1 $ implies that node-$k$ is offloading the data at slot-$n$, and vice versa. During this period of time, we assume the offloaded data amount is $ l_{n,k} $, then the ground-air transmission is required to satisfy
\begin{equation} \label{eq:l_nk} % equ1 \ref{equ1} 
{\chi _{n,k}}{\rho _n}\frac{T}{N}{B_1}{\log _2}\left( {1 + \frac{{{p_{n,k}}{h_{n,k}}}}{{{\sigma_0 ^2}}}} \right) \ge {l_{n,k}}, \:\: \forall k \in \mathcal{K},\forall n \in \mathcal{N},
\end{equation}
where $ B_1 $ is the ground-air communication bandwidth, $ h_{n,k} $ is the channel condition in~(\ref{eq:h_nk}), $ \sigma_0^2 $ is the noise power, and $ p_{n,k} $ is the allocated transmit power constrained by
\begin{equation} \label{eq:pwr_con} % equ1 \ref{equ1} 
0 \le {p_{n,k}} \le p_k^{\max }, \quad \forall k \in \mathcal{K},\forall n \in \mathcal{N},
\end{equation}
with $p_k^{\max }$ being the power budget at node-$k$. Considering the data offloading for each node during the whole time horizon, the offloading needs to satisfy the required data amount, given as
\begin{equation} \label{eq:amt_con} % equ1 \ref{equ1} 
\sum\limits_{n = 1}^N {{l_{n,k}} \ge {{D}_k}}, \quad\forall k \in \mathcal{K},
\end{equation}
where $ D_k $ refers to the amount of data that each node-$k$ needs to offload during the period $T$. For the data computation, the UAV as an edge server processes the collected data from all ground nodes, which satisfies the computation capacity at the UAV, given as
\begin{equation} \label{eq:comp_con} % equ1 \ref{equ1} 
{c_0}\sum\limits_{k = 1}^K {{l_{n,k}} \le } \rho_n\frac{T}{N}{f^{\max }}, \quad\forall n \in \mathcal{N},
\end{equation}
where $ c_0 $ and $ f^{\max} $ are the computation capability in cycle per bit and the highest computation frequency, respectively, of the UAV. Finally, the computation results are forwarded to the satellite from the UAV through the air-space link. For the results, we assume that data size of the results is proportional to the incoming computation data, with a coefficient given by $ o \in (0,1) $, depending on the computation types. Thus, during time slot-$n$ with a time duration of $ (1-\rho_n)T/N $ for results forwarding, the transmitted data amount is $ o\sum\nolimits_{k = 1}^K {{l_{n,k}}} $. Meanwhile, in the $n$-th time slot, the transmit power of the single-antenna UAV is denoted by $ p_n^{\text{(UAV)}} $, constrained by $ p^{\text{(UAV)},\max} $, and the receive beamforming at the $M$-antenna satellite is denoted by $ \bm{\omega}_n^{\text{(sat)}} $. Since the receive beamforming is of unit modulus, we introduce a new variable $ \bm{w}_n $ in slot-$n$ with its square modulus and normalized direction corresponding to the UAV transmit power and satellite receive beamforming, respectively. In this respect, we have simplified notations $ \left\{\bm{w}_n\right\}_{n\in\mathcal{N}} $ with the constraints that
\begin{equation} \label{eq:w_con}
\left\| \bm{w}_n \right\|^2 \le p^{\text{(UAV)},\max}, \quad \forall n\in\mathcal{N}.
\end{equation}
Similarly, along with the air-space channel in~(\ref{eq:g_n}), the results forwarding is subject to the constraints
\begin{equation} \label{eq:result_con} % equ1 \ref{equ1} 
\left( {1 - {\rho _n}} \right)\frac{T}{N}{B_2}{\log _2}\left( 1 + \left\| \bm{g}_n^H \bm{w}_n \right\|^2  \right) \ge o\sum\limits_{k = 1}^K {{l_{n,k}}}, \:\: \forall n \in \mathcal{N},
\end{equation}
to ensure the transmitted amount of results, where the noise power has been normalized in the UAV-satellite links. With this constraint, we ensure that the UAV can successfully forward computed results to the satellite. In this regard, it needs to balance the time allocation, task offloading, and link resources to guarantee the smoothness in the computation and forwarding processes, which is crucial for service provisioning in the NTN environments.

\subsection{Energy Consumption}
Along with the NTN-enabled integrated communication and computation process detailed above, the energy consumed by the parties involved in the system is analyzed as follows. Specifically, the ground node is active when transmitting their collected data to the UAV in the offloading phase, where the energy consumption is given as
\begin{equation} \label{eq:E_1} % equ1 \ref{equ1} 
E_1 = \sum\limits_{n = 1}^N {\rho _n}{\frac{T}{N}} \sum\limits_{k = 1}^K {\chi _{n,k}}{{p_{n,k}}} ,
\end{equation}
counting all the nodes in the whole time horizon. For the data computation at the UAV, the energy consumption is determined by the computed data amount and allowed time, where the computation frequency can be dynamically adapted for energy conservation. In this regard, the computation energy at the UAV is specified as
\begin{equation} \label{eq:E_2} % equ1 \ref{equ1} 
E_2 = \sum\limits_{n = 1}^N {\frac{{\gamma {{\left( {{c_0}\sum\limits_{k = 1}^K {{l_{n,k}}} } \right)}^3}}}{{{{\left( {\frac{T}{N}{\rho _n}} \right)}^2}}}},
\end{equation}
where $\gamma $ is the effective capacitance coefficient of UAV edge. For the results forwarding from the UAV to the satellite, the energy consumption is determined by the power allocation at UAV, given as
\begin{equation} \label{eq:E_3} % equ1 \ref{equ1} 
E_3 = \sum\limits_{n = 1}^N {{{\left\| {\bm{w}_n} \right\|}^2}\left( {1 - {\rho _n}} \right)}\frac{T}{N}.
\end{equation}
Moreover, since the UAV flies to visit the ground nodes to collect the offloaded data, additional energy consumption for the propulsion is required, given as
\begin{equation} \label{eq:E_prop} % equ1 \ref{equ1} 
\begin{aligned}
P\left(V_n\right) =&\:P_{0}\left(1+\frac{3V_n^{2}}{U_{\text{tip}}^{2}}\right)+P_{i}\left(\sqrt{1+\frac{V_n^{4}}{4v_{0}^{4}}}-\frac{V_n^{2}}{2v_{0}^{2}}\right)^{1/2} \\
&+\frac{1}{2}d_{0}\varpi sAV_n^{3},
\end{aligned}
\end{equation}
for time slot-$n$ with flying speed $ V_n =  \left\| {{\bm{q}_n} - {\bm{q}_{n - 1}}} \right\| / T/N $, where $ P_0 $ and $ P_i $ are constants representing the blade profile power and induced power while hovering, $ U_{\text{tip}} $ is the tip speed of the rotor blade, $ v_0 $ is the mean rotor induced velocity in hover, $ d_0 $ and $ s $ are the fuselage drag ratio and rotor solidity, respectively, and $ \varpi $ and $ A $ denote the air density an rotor disc area, respectively~\cite{cja}. Here we adopt the rotate-wing UAV model for more flexible flying services. As such, the propulsion energy of the UAV is obtained as
\begin{equation} \label{eq:E_4} % equ1 \ref{equ1} 
E_4 = \sum\limits_{n = 1}^N {{P\left(V_n\right)}\frac{T}{N}}.
\end{equation}
Given the on-board battery of the UAV to support the flying with propulsion energy capacity denoted by $ E^{\text{(prop)}} $, then UAV flying is subject to the constraint that
\begin{equation} \label{eq:prop_con}
	\sum\limits_{n = 1}^N {{P\left(V_n\right)}\frac{T}{N}} \le E^{\text{(prop)}}.
\end{equation}

Therefore, with the energy model for each part of the proposed NTN-enabled integrated communication and computation, the overall energy consumption is obtained as
\begin{equation} \label{eq:E}
	E = E_1 + \eta_1(E_2 + E_3) + \eta_2 E_4,
\end{equation}
where $ \eta_1 $ and $ \eta_2 $ are the weighting factors to distinguish different energy sources and usage. Also, the weighting factor allows higher compatibility to incorporate UAVs of different types and parameters for more flexible models.

\subsection{Uncertainty}

There are generally a wide range of uncertainties existed in wireless systems. For our considered network, the motion of UAV and satellite induces significant large-scale channel variations. Although we can approximate the UAV-satellite distance through the positioning systems, the exact location of the non-terrestrial nodes is still rather difficult to obtain. Correspondingly, the uncertainties associated with the phase effect of the UAV-satellite link originated from the large-scale network dynamics need to be explicitly addressed. Particularly, the phase in time slot-$n$ can be specified as
\begin{equation} \label{eq:theta} % equ1 \ref{equ1} 
{\bm{\theta} _n} = {\hat{\bm{\theta}} _n} + {\tilde{\bm{\theta}}_n},\quad\forall n \in \mathcal{N},
\end{equation}
where $ \hat{\bm{\theta}}_n $ is the estimation and $ {\tilde{\bm{\theta}}_n} $ is the associated error. The estimation of the phase can be obtained based on the space geometry information as $ \hat{\bm{\theta}}_n = \left[ \hat{{\theta}}_{n,m} \right]_{m=1,2,\cdots,M} $ with
\begin{equation} \label{eq:theta_nm} % equ1 \ref{equ1} 
\theta_{n,m} =  { - \frac{{2\pi {d^{\left(\text{SU}\right)}_n}}}{\lambda } - \frac{{2\pi d}}{\lambda }\left( {m-1} \right){\phi^\text{(SU)}}} ,\:\: m =1,2,\cdots M,
\end{equation}
where $d$ is the satellite antenna separation and $ \phi^\text{(SU)} $ is the cosine of elevation angle between the link and antenna direction. The associated error has many potential contributing factors, including the inaccuracy of the locations of UAV and satellite, the imperfection of the circuit components such as the oscillator and downconverter, etc. Therefore, it can be quite difficult to describe the uncertainty with conventional bound information. Rather, considering the varieties of the error sources, we exploit the central-limit principle and employ the Gaussian distribution to characterize the error as
\begin{equation} \label{eq:} % equ1 \ref{equ1} 
{\tilde{\bm{\theta}}_n} \sim\mathscr{N}\left(\bm{0}, \varrho^2\bm{I}_M\right), \quad\forall n \in \mathcal{N}, 
\end{equation}
where $ \varrho^2 $ is the variance of phase error, assumed to be identical at all antennas and can be obtained through training or historical observations.

In accordance with the phase uncertainties, the UAV-satellite link becomes random, and thus the results forwarding constraint in~(\ref{eq:result_con}) can no longer be settled in a deterministic manner. Instead, we adopt the chance constraints as
\begin{equation} \label{eq:chance_con} % equ1 \ref{equ1}
\begin{aligned}
& \mathbb{P}_{{\tilde{\bm{\theta}}_n}} \left\{ \left( {1 - {\rho _n}} \right)\frac{T}{N}{B_2}{\log _2}\left( 1 + \left\| \bm{g}_n^H \bm{w}_n \right\|^2  \right) \ge o\sum\limits_{k = 1}^K {{l_{n,k}}} \right\} \\
 \geq \: & 1- {\epsilon}_n,\quad\forall n\in\mathcal{N}, 
\end{aligned}
\end{equation}
where ${\epsilon_n}$ is the threshold of the probability that the result forwarding fails the required amount during time slot-$n$.

\subsection{Problem}

Based on the discussions above, we incorporate the energy consumption during the offloading, computation, and forwarding processes concerning different roles in the system, and arrive at the energy optimization problem as
\begin{IEEEeqnarray}{CL} \label{eq:problem}
	\IEEEyesnumber \IEEEnosubnumber*
	\min_{\substack{ \left\{ {{\bm{q}_n},{\rho_n},{\bm{w}_n}} \right\}_{\forall n \in \mathcal{N}} \\ \left\{ {{p_{n,k}},{\chi _{n,k}},{l_{n,k}}} \right\}_{ \forall n \in \mathcal{N}, \forall k \in \mathcal{K} } }} & E \\
	{\rm{s.t.}} & (\ref{eq:q_sf}), (\ref{eq:speed}), \nonumber \\
	& (\ref{eq:chi_con}), (\ref{eq:pwr_con}), (\ref{eq:w_con}), (\ref{eq:prop_con}), \nonumber \\
	& (\ref{eq:l_nk}), (\ref{eq:amt_con}), (\ref{eq:comp_con}), (\ref{eq:chance_con}), \nonumber \IEEEeqnarraynumspace
\end{IEEEeqnarray}
which considers the ground nodes scheduling and offloading, UAV flying, computation, and forwarding, in the presence of UAV-satellite link uncertainties. Although the formulated problem emphasizes the energy issue, the latency is also implicitly incorporated as we can flexibly adjust the overall allowed time horizon to fit the service requirement. Also, the slot-wise time portion allocation also presents the latency requirements for the transmissions in both the ground-UAV offloading and the UAV-satellite forwarding stages.

The formulated problem is rather intricate since it physically interprets a complicated integrated communication and computation process, and mathematically has series of variables coupled in complex constraints. Technically, the difficulties mainly include three folds. First, the problem concerns both the slotted transmission behaviors under given topology as well as the flying trajectory along the whole time horizon. Second, the optimization incorporates both continuous and integral variables, interplaying with each other and hindering efficient solutions. Third, there are constraints in probabilistic forms along with the deterministic ones, requiring additional procedures to tackle towards effective algorithm design.

\section{Proposed Solution} \label{sec5}

In this section, the formulated problem is analyzed and decomposed, with subproblems solved effectively. Then, the algorithm is proposed to achieve energy-efficient integrated communication and computation.

\subsection{Reformulation and Decomposition}

For the formulated problem in~(\ref{eq:problem}), we recast the beamforming vector in the form of semidefinite programming. In this respect, we introduce $ \left\{\bm{W}_n\right\}_{n\in\mathcal{N}} $ with $ \bm{W}_n = \bm{w}_n\bm{w}_n^H $, and thus the power constraints associated with UAV beamforming is obtained as
\begin{equation} \label{eq:W_con} \left\{
	\begin{aligned}
		&\mathsf{tr}(\bm{W}_n) \le p^{\text{(UAV)},\max}, \\
		&\mathsf{rank}(\bm{W}_n) = 1, \quad \bm{W}_n \succeq \bm{0},
	\end{aligned} \quad \forall n \in\mathcal{N}. \right.
\end{equation}
Meanwhile, the probabilistic-form constraints in~(\ref{eq:chance_con}) need to be converted into deterministic forms to facilitate the problem solving. As such, we first separate the variables associated with uncertainties form the rest. Due to the angular information errors in~(\ref{eq:theta}), the UAV-satellite link also can be decomposed into the deterministic part and uncertain part. In this respect, by introducing $ {{{\hat{\bm{g}}}_n}} = \sqrt {\frac{{{G_{n}^{\text{(PL)}}}{G_{t}}{G_{r}}}}{{\kappa B_2t}}} \bm{r}_n^{-\frac{1}{2}} \odot {e^{j{\hat{\bm{\theta}}_n }}} $ denoting the deterministic part based on estimated phases, the chance constraints in~(\ref{eq:chance_con}) can be reformulated as
\begin{equation} \label{eq:chance_con2}
\begin{aligned} % equ1 \ref{equ1} 
& \mathbb{P}_{{\tilde{\bm{\theta}}_n}} \left\{{{({e^{j{\tilde{\bm{\theta}}_n}}})^H}\mathsf{diag}^H\left( { {{\hat{\bm{g}}_n}} } \right){\bm{W}_n}\mathsf{diag}\left( { {{\hat{\bm{g}}_n}} } \right){e^{j{\tilde{\bm{\theta}}_n}}}} \right. \\
& \qquad \quad \left. -{2^{\frac {o\sum\nolimits_{k = 1}^K {{l_{n,k}}}}{\frac{T}{N}\left( {1 - {\rho _n}} \right){B_2}}}+1}\ge 0\right\} \geq 1- {\epsilon}_n, \quad\forall n\in\mathcal{N},
\end{aligned} 
\end{equation}
by rearranging the terms therein. Then, the left-hand side of the condition inequality in~(\ref{eq:chance_con2}) appears in the quadratic form of the exponential function in terms of angular uncertainties. By leveraging~\cite[Lemma~1]{apprx1}, we introduce $ \left\{\bm{Z}_n\right\}_{n\in\mathcal{N}} $ with $ \bm{Z}_n = \mathsf{diag}^H\left( { {{\hat{\bm{g}}_n}} } \right){\bm{W}_n}\mathsf{diag}\left( { {{\hat{\bm{g}}_n}} } \right) $, and arrive at the approximation that
\begin{equation}
\begin{aligned}
& \: \exp^H(j{\tilde{\bm{\theta}}_n})\bm{Z}_n\exp(j{\tilde{\bm{\theta}}_n}) \\
\approx & \: {\tilde{\bm{\theta}}_n}^Tf(\bm{X}_n){\tilde{\bm{\theta}}_n} + {\tilde{\bm{\theta}}_n}^T g(\bm{Y}_n)+\sum_{i,j}[\bm{Z}_n]_{i,j},
\end{aligned}
\end{equation}
where $\bm{X}_n=\mathsf{Re}\{\bm{Z}_n\}$, $\bm{Y}_n=\mathsf{Im}\{\bm{Z}_n\}$ and
\begin{align}
&{\left[ {f\left( \bm{X}_n \right)} \right]_{i,j}} = 
\begin{cases}
{[\bm{X}_n]_{i,j} - \sum\limits_{j^\prime} {[\bm{X}_n]_{i,j^\prime}},} & \quad \text{if } i = j, \\
{[\bm{X}_n]_{i,j},} & \quad \text{if } i \neq j,
\end{cases} \\
&{\left[ {g\left( \bm{Y}_n \right)} \right]_{i,1}} = 2\sum\limits_{j^\prime} {{{\left[ \bm{Y}_n \right]}_{i,j^\prime}}}.
\end{align}
Then, the chance constraints with respect the angular uncertainties in~(\ref{eq:chance_con2}) are represented as
\begin{equation} \label{eq:chance_con3}
\begin{aligned}
& \mathbb{P}_{{\tilde{\bm{\theta}}_n}} \left\{ {\tilde{\bm{\theta}}_n}^Tf(\bm{X}_n){\tilde{\bm{\theta}}_n} + {\tilde{\bm{\theta}}_n}^T g(\bm{Y}_n)+\sum_{i,j}[\bm{Z}_n]_{i,j} \right. \\
& \qquad \quad \left. -{2^{\frac {o\sum\nolimits_{k = 1}^K {{l_{n,k}}}}{\frac{T}{N}\left( {1 - {\rho _n}} \right){B_2}}}+1}\ge 0\right\} \geq 1- {\epsilon}_n, \quad\forall n\in\mathcal{N},
\end{aligned} 
\end{equation}
where the left-hand side of the condition inequality now becomes the quadratic of the uncertainties. Given the Gaussian distribution of the uncertainties which induces the probabilistic constraints in~(\ref{eq:chance_con3}), we can exploit the Bernstein-Type inequality to tackle it effectively~\cite{apprx2}. Specifically, the Bernstein-Type inequality-induced sufficient condition for~(\ref{eq:chance_con3}) is given as
\begin{subnumcases}{\label{eq:BTI}}
\begin{aligned}
\mathsf{tr}(\bm{Q}_n) - 2\sqrt{-\ln(\epsilon_n)}x_n + 2\ln(\epsilon_n)y_n + s_n \geq 0,\\
\forall n\in \mathcal{N},
\end{aligned} \label{eq:BTI_a}\\
\left\| \begin{bmatrix}
\mathsf{vec}(\bm{Q}_n) \\
\sqrt{2}\bm{r}_n
\end{bmatrix} \right\|_2 \leq x_n , \quad\forall n\in \mathcal{N}, \label{eq:BTI_b}\\
y_n\bm{I}_{M} + \bm{Q}_n \succeq \bm{0}, \quad\forall n\in \mathcal{N}, \label{eq:BTI_c}
\end{subnumcases}
where the non-negative $ \left\{x_n\right\}_{n\in\mathcal{N}} $ and $ \left\{y_n\right\}_{n\in\mathcal{N}} $ are introduced auxiliaries and
\begin{subnumcases}
	\bm{Q}_n=\xi^2{f(\bm{X}_n)}, \quad\forall n\in \mathcal{N}, \\
	\bm{r}_n=\frac{1}{2}\xi{g(\bm{Y}_n)}, \quad\forall n\in \mathcal{N}, \\
	s_n=\sum_{i,j}[\bm{Z}_n]_{i,j}-{2^{\frac {o\sum\nolimits_{k = 1}^K {{l_{n,k}}}}{\frac{T}{N}\left( {1 - {\rho _n}} \right){B_2}}}+1}, \quad\forall n\in \mathcal{N}.
\end{subnumcases}
Therefore, with semidefinite programming and the treatment of the chance constraints above, we achieve the reformulation of problem in~(\ref{eq:problem}) as
\begin{IEEEeqnarray}{CL} \label{eq:problem_equiv}
	\IEEEyesnumber \IEEEnosubnumber*
	\min_{\substack{ \left\{ {{\bm{q}_n},{\rho_n},{\bm{W}_n}} \right\}_{\forall n \in \mathcal{N}} \\ \left\{ {{p_{n,k}},{\chi _{n,k}},{l_{n,k}}} \right\}_{ \forall n \in \mathcal{N}, \forall k \in \mathcal{K} } \\ \left\{x_n, y_n\right\}_{\forall n\in\mathcal{N}} }} & E \\
	{\rm{s.t.}} & (\ref{eq:q_sf}), (\ref{eq:speed}), \nonumber \\
	& (\ref{eq:chi_con}), (\ref{eq:pwr_con}), (\ref{eq:W_con}), (\ref{eq:prop_con}), \nonumber \\
	& (\ref{eq:l_nk}), (\ref{eq:amt_con}), (\ref{eq:comp_con}), (\ref{eq:BTI}). \nonumber \IEEEeqnarraynumspace
\end{IEEEeqnarray}
Compared with the original equivalence, the reformulated problem above now has all constraints in a deterministic form and thus facilitates the analysis.

Further, by analyzing the physical interactions during the communication and computation processes and the mathematical forms indicated in the objective function and constraints, we decompose the problem into four subproblems as follows. The first problem concerns time-slotted offloading data amount and user scheduling, given as
\begin{IEEEeqnarray}{CL} \label{eq:problem_1}
	\IEEEyesnumber \IEEEnosubnumber*
	\min_{\substack{  \left\{ {\chi_{n,k}},{l_{n,k}} \right\}_{ \forall n \in \mathcal{N}, \forall k \in \mathcal{K} } \\ \left\{x_n, y_n\right\}_{\forall n\in\mathcal{N}} }} & E_1 + \eta_1E_2 \\
	{\rm{s.t.}} &  (\ref{eq:chi_con}), (\ref{eq:l_nk}), (\ref{eq:amt_con}), (\ref{eq:comp_con}), (\ref{eq:BTI}). \nonumber
	\IEEEeqnarraynumspace
\end{IEEEeqnarray}
The second subproblem decides the portion of time in each slot used for the ground node-UAV transmissions and UAV-satellite transmissions, specified as
\begin{IEEEeqnarray}{CL} \label{eq:problem_2}
	\IEEEyesnumber \IEEEnosubnumber*
	\min_{  \left\{\rho_n,x_n, y_n\right\}_{\forall n\in\mathcal{N}} } & E_1 + \eta_1E_3 \\
	{\rm{s.t.}} &  (\ref{eq:l_nk}), (\ref{eq:comp_con}), (\ref{eq:BTI}). \nonumber 
	\IEEEeqnarraynumspace\IEEEeqnarraynumspace
\end{IEEEeqnarray}
The third subproblem is about the transmission beamforming for the result forwarding from the UAV to the satellite, represented as
\begin{IEEEeqnarray}{CL} \label{eq:problem_3}
	\IEEEyesnumber \IEEEnosubnumber*
	\min_{ \left\{\bm{W}_n,x_n, y_n\right\}_{\forall n\in\mathcal{N}} } & \eta_1E_3 \\
	{\rm{s.t.}} & (\ref{eq:W_con}), (\ref{eq:BTI}). \nonumber
	\IEEEeqnarraynumspace\IEEEeqnarraynumspace\IEEEeqnarraynumspace
\end{IEEEeqnarray}
The fourth subproblem incorporates the UAV trajectory design as well as the ground node transmit power optimization during the flight, cast as
\begin{IEEEeqnarray}{CL} \label{eq:problem_4}
	\IEEEyesnumber \IEEEnosubnumber*
	\min_{\substack{ \left\{ {\bm{q}_n} \right\}_{\forall n \in \mathcal{N}} \\ \left\{ {p_{n,k}} \right\}_{ \forall n \in \mathcal{N}, \forall k \in \mathcal{K} } }} & E_1 + \eta_2 E_4 \\
	{\rm{s.t.}} & (\ref{eq:q_sf}), (\ref{eq:speed}), (\ref{eq:prop_con}), \nonumber \\
	&  (\ref{eq:l_nk}), (\ref{eq:pwr_con}) . \nonumber
	\IEEEeqnarraynumspace\IEEEeqnarraynumspace\IEEEeqnarraynumspace\IEEEeqnarraynumspace
\end{IEEEeqnarray}
The decomposition above is based on the three-stage communication and computation process, where, basically, the first two subproblems track the behaviors between ground nodes and UAV, the third subproblem tackles the actions between the UAV and satellite, and the last subproblem addresses the UAV flight during the whole procedure.

\subsection{Analysis and Algorithm}

In this section, we analyze and solve the decomposed subproblems, addressing the offloading, scheduling, transmission, and trajectory, respectively, which are then synthesized under the block coordinate descent framework to achieve the solution to the original problem.

For the first subproblem in~(\ref{eq:problem_1}), it consists the integral optimization variables $ \left\{\chi_{n,k}\right\}_{n\in\mathcal{N},k\in\mathcal{K}} $ to indicating whether ground node-$k$ is scheduled in time slot-$n$. Given the evident difficulties in handling the integral variables directly, we relax it as a continuous variable in the interval $ [0,1] $. Then, we can conveniently verify that the relaxed problem is convex, as the objective is a convex function and the constraints are linear functions with respect to the optimization variables, with arrangement of the terms. The convex problem can be solved efficiently with off-the-shelf tools, where the integral optimization variables can be restored by rounding the obtained results.

For the second subproblem in~(\ref{eq:problem_2}), the objective function is linear to the optimization variable while the constraints in~(\ref{eq:l_nk}) and (\ref{eq:comp_con}) are also linear, where the non-convexity lines in the Bernstein-type inequality induced constraints in~(\ref{eq:BTI}). In this respect, we introduce $ \left\{\psi_n\right\}_{n\in\mathcal{N}} $ such that
\begin{equation}  \label{eq:psi}
{\psi _n} \ge{2^{\frac {o\sum\nolimits_{k = 1}^K {{l_{n,k}}}}{\frac{T}{N}\left( {1 - {\rho _n}} \right){B_2}}}-1}, \quad \forall n \in \mathcal{N},
\end{equation}
which induces that
\begin{equation} \label{eq:BTI_a_relaxed} % equ1 \ref{equ1} 
\begin{aligned}
&\mathsf{tr}(\bm{Q}_n)-2\sqrt{-\ln(\epsilon_n)}x_n+2\ln(\epsilon_n)y_n \\
&\qquad\qquad + \sum_{i,j}{[\bm{Z}_n}]_{i,j}-{\psi _n} \geq 0, \quad\forall n\in\mathcal{N}.
\end{aligned}
\end{equation}
Further, the condition in~(\ref{eq:psi}) can be recast as
\begin{equation} \label{eq:psi_con} % equ1 \ref{equ1} 
{\log _2}\left( {{\psi _n} + 1} \right) \ge \frac{{o\sum\nolimits_{k = 1}^K {{l_{n,k}}} }}{{\frac{T}{N}\left( {1 - {\rho _n}} \right){B_2}}}, \quad \forall n \in \mathcal{N},
\end{equation}
where the non-convexity lies on the right-hand side with respect to $ \left\{ \rho_n \right\}_{n\in\mathcal{N}} $. In this respect, we can employ the successive convex approximation that
\begin{equation} \label{eq:rho_sca} % equ1 \ref{equ1} 
\frac{1}{{1 - {\rho _n}}} \le \frac{ 1 + \rho_n - 2\rho_n^\circ }{{\left( {1 - \rho _n^ \circ } \right)}^2} ,\quad\forall n \in \mathcal{N},
\end{equation}
by using the first-order Taylor expansion at $ \left\{ \rho_n^\circ \right\}_{n\in\mathcal{N}} $. Then, the constraint in~(\ref{eq:psi_con}) is approximated by using the upper-bound counterpart of its right-hand side such that
\begin{equation} \label{eq:psi_sca} % equ1 \ref{equ1} 
{\log _2}\left( {{\psi _n} + 1} \right) \ge \frac{{o\sum\nolimits_{k = 1}^K {{l_{n,k}}} }}{{\frac{T}{N}{B_2}}} \cdot \frac{ 1 + \rho_n - 2\rho_n^\circ }{{\left( {1 - \rho _n^ \circ } \right)}^2}  , \quad \forall n \in \mathcal{N},
\end{equation}
which becomes convex with respect to the optimization variables. Based on the discussions above, we now arrive at the convex counterpart of the problem in~(\ref{eq:problem_2}) as
\begin{IEEEeqnarray}{CL} \label{eq:problem_2_sca}
	\IEEEyesnumber \IEEEnosubnumber*
	\min_{  \left\{\rho_n,x_n, y_n, \psi_n\right\}_{\forall n\in\mathcal{N}} } & E_1 + \eta_1E_3 \\
	{\rm{s.t.}} &  (\ref{eq:l_nk}), (\ref{eq:comp_con}), \nonumber \\ 
	& (\ref{eq:BTI_a_relaxed}), (\ref{eq:BTI_b}), (\ref{eq:BTI_c}), (\ref{eq:psi_sca}), \nonumber 
	\IEEEeqnarraynumspace
\end{IEEEeqnarray}
which is approximated at $ \left\{ \rho_n^\circ \right\}_{n\in\mathcal{N}} $. In this respect, we can solve a series of convex problems in the form of~(\ref{eq:problem_2_sca}), with the obtained optimum to update the approximation point, and thus approach the solution to the original problem in~(\ref{eq:problem_2}).

For the third problem to tackle the UAV-satellite transmission beamforming, it is evident that the non-convexity is due to the rank-one constraint. In this respect, given the non-negative definiteness of the $ \bm{W}_n $, $ \forall n\in\mathcal{N} $, then we have that $ \mathsf{tr}\left(\bm{W}_n\right) - \lambda_{\max}\left(\bm{W}_n\right) \ge 0 $. To reach the condition that $ \mathsf{tr}\left(\bm{W}_n\right) - \lambda_{\max}\left(\bm{W}_n\right) = 0 $, which indicates that $ \bm{W}_n $ has one single non-zero eigenvalue and thus guarantees the rank-one constraint, we simply additionally require that $ \mathsf{tr}\left(\bm{W}_n\right) - \lambda_{\max}\left(\bm{W}_n\right) \le 0 $. We then use the newly induced condition as a penalty function and thus update the objective function of~(\ref{eq:problem_3}) as
\begin{equation}
	\min_{ \left\{\bm{W}_n,x_n, y_n\right\}_{n\in\mathcal{N}} } \eta_1\left(E_3 + \sum\limits_{n=1}^N \zeta_n \left( \mathsf{tr}\left(\bm{W}_n\right) - \lambda_{\max}\left(\bm{W}_n\right) \right)\right),
\end{equation}
where $ \zeta_n \ge 0$, $ \forall n\in\mathcal{N} $ is the weight coefficients large enough to force $ \mathsf{tr}\left(\bm{W}_n\right) - \lambda_{\max}\left(\bm{W}_n\right) $ to approach zero. Since the maximum eigenvalue function in the penalized objective is convex, we employ the sub-gradient of the maximum eigenvalue function that $ \partial \lambda_{\max} \left(\bm{W}_n\right) = \bm{w}_n^{\max} \left(\bm{w}_n^{\max}\right)^H $ with $ \bm{w}_n^{\max} $ being unit-norm eigenvector corresponding to the maximum eigenvalue. Then, we exploit the approximation that
\begin{equation} \label{eq:lambda_sca}
\begin{aligned}
	\lambda_{\max}\left(\bm{W}_n\right) \ge \lambda_{\max}\left(\bm{W}_n^\circ\right) + \left\langle \bm{w}_n^{\max,\circ} \left(\bm{w}_n^{\max,\circ}\right)^H, \right. \\
	\left. \bm{W}_n -\bm{W}_n^\circ \right\rangle,  \forall n\in\mathcal{N},
\end{aligned}
\end{equation}
which is expanded at $ \left\{\bm{W}_n^\circ\right\}_{n\in\mathcal{N}} $ with counterpart eigenvector $ \left\{\bm{w}_n^{\max,\circ}\right\}_{n\in\mathcal{N}} $ used in the subgradient. By using the right-hand side of~(\ref{eq:lambda_sca}) to approximate the maximum eigenvalue function in the objective and neglecting the constant terms, we have the optimization that
\begin{IEEEeqnarray}{CL} \label{eq:problem_3_sca}
	\IEEEyesnumber \IEEEnosubnumber*
	\min_{ \left\{\bm{W}_n,x_n, y_n\right\}_{\forall n\in\mathcal{N}} } & 
	\begin{aligned}
		 \eta_1E_3 & + \eta_1\sum\limits_{n=1}^N \zeta_n \left( \mathsf{tr}\left(\bm{W}_n\right) \right. \\
		& \left. - \left\langle \bm{w}_n^{\max,\circ} \left(\bm{w}_n^{\max,\circ}\right)^H, \bm{W}_n \right\rangle \right)
	\end{aligned} \\
	{\rm{s.t.}} & (\ref{eq:W_con}), (\ref{eq:BTI}), \nonumber
	\IEEEeqnarraynumspace\IEEEeqnarraynumspace\IEEEeqnarraynumspace
\end{IEEEeqnarray}
as approximated at $ \left\{\bm{W}_n^\circ\right\}_{n\in\mathcal{N}} $ with the rank-one constraint in~(\ref{eq:W_con}) being removed. Then, we can solve a series of problems in the form of~(\ref{eq:problem_3_sca}) with the approximation point updated based on the obtained optimum, and finally achieve the solution to the UAV-satellite beamforming problem in~(\ref{eq:problem_3}).

For the fourth subproblem concerns the UAV trajectory along with the ground node transmission power strategy, as specified in~(\ref{eq:problem_4}), we can readily see the transmit power related objective and constraints all convex, and thus we need to tackle the non-convexity associated with the trajectory design. In this regard, the non-convexity originates from the complicated propulsion energy against flying speed, affecting both the objective and the constraints. Thus, we introduce a new variable $ \tilde E_4 \ge 0 $ such that $ \sum\nolimits_{n = 1}^N {{P\left(V_n\right)}\frac{T}{N}} \le \tilde E_4 $, and reformulate the problem in~(\ref{eq:problem_4}) as
\begin{IEEEeqnarray}{CL} \label{eq:problem_4_eqiv}
	\IEEEyesnumber \IEEEnosubnumber*
	\min_{\substack{ \tilde E_4, \left\{ {\bm{q}_n} \right\}_{\forall n \in \mathcal{N}} \\ \left\{ {p_{n,k}} \right\}_{ \forall n \in \mathcal{N}, \forall k \in \mathcal{K} } }} & E_1 + \eta_2 \tilde E_4 \\
	{\rm{s.t.}} & (\ref{eq:q_sf}), (\ref{eq:speed}), (\ref{eq:l_nk}), (\ref{eq:pwr_con}), \nonumber \\
	&  \sum\limits_{n = 1}^N {{P\left(V_n\right)}\frac{T}{N}} \le \min\left\{ \tilde E_4, E^{\text{(prop)}} \right\} .\label{eq:prop_con_2}
	\IEEEeqnarraynumspace
\end{IEEEeqnarray}
For notation simplicity, we denote $ \Delta_n = \left\| \bm{q}_n - \bm{q}_{n-1} \right\| $ and thus $ V_n = \Delta_n/T/N $, $ \forall n\in\mathcal{N} $, then the constraint in~(\ref{eq:prop_con_2}) is rewritten as
\begin{equation} \label{eq:prop_con_3} % equ1 \ref{equ1} 
\begin{aligned}
& \sum_{n=1}^N P_0\left(\frac{T}{N}+\frac{3\Delta_n^2}{U_{\text{tip}}^2{T/N}}\right) + P_i \left(\sqrt{\left(\frac{T}{N}\right)^4 + \frac{\Delta_n^4}{4v_0^4}} \right. \\
& \quad \left. - \frac{\Delta_n^2}{2v_0^2}\right)^{1/2} + \frac{1}{2}d_0\rho sA\frac{\Delta_n^3}{\left(T/N\right)^2}\leq \min\left\{ \tilde E_4, E^{\text{(prop)}} \right\}.
\end{aligned}
\end{equation}
Given the complicated expression above, we introduce $ \left\{z_n\right\}_{n\in\mathcal{N}} \ge 0 $ such that
\begin{equation} \label{eq:z_n} % equ1 \ref{equ1} 
z_n^2 \geq \sqrt{\left(\frac{T}{N}\right)^4+\frac{\Delta_n^4}{4v_0^4}}-\frac{\Delta_n^2}{2v_0^2}, \quad \forall n\in\mathcal{N},
\end{equation}
with the condition in~(\ref{eq:prop_con_3}) reinterpreted as
\begin{equation} \label{eq:prop_con_4} % equ1 \ref{equ1} 
\begin{aligned}
 \sum_{n=1}^N P_0\left(\frac{T}{N}+\frac{3\Delta_n^2}{U_{\text{tip}}^2{T/N}}\right) + P_i z_n + \frac{1}{2}d_0\rho sA\frac{\Delta_n^3}{\left(T/N\right)^2} \\
 \leq \min\left\{ \tilde E_4, E^{\text{(prop)}} \right\}.
\end{aligned}
\end{equation}
Then, the constraint in~(\ref{eq:prop_con_4}) is jointly convex with respect to $ \left\{ \bm{q}_n, z_n\right\}_{n\in\mathcal{N}} $ and $ \tilde{E}_4 $. While for the non-convexity remained in~(\ref{eq:z_n}), we rearrange the terms therein and have that
\begin{equation} \label{eq:z_n_2} % equ1 \ref{equ1} 
z_n^2 + \frac{\Delta_n^2}{v_0^2} \geq \frac{\left(T/N\right)^4}{z_n^2}, \quad \forall n\in\mathcal{N}.
\end{equation}
Since the left-hand side of the inequalities above is convex, we can exploit the first-order expansion at $ \left\{ \bm{q}_n^\circ, z_n^\circ\right\}_{n\in\mathcal{N}} $ and arrive at
\begin{equation} \label{eq:z_n_sca}
\begin{aligned}
z_n^2+\frac{\Delta_n^2}{v_0^2} \ge & \left(z_n^\circ\right)^2 + 2z_n^\circ\left(z_n-z_n^\circ\right)-\frac{\left\|\bm{q}_{n+1}^\circ-\bm{q}_n^\circ\right\|^2}{v_0^2} \\
& +\frac{2}{v_0^2}\left[\bm{q}_{n+1}^\circ-\bm{q}_n^\circ\right]^T\left[\bm{q}_{n+1}-\bm{q}_n\right], \quad \forall n\in\mathcal{N}.
\end{aligned}
\end{equation}
We use the lower-bound approximation to replace the counterpart in~(\ref{eq:z_n_2}) and thus have the following convex constraint that
\begin{equation} \label{eq:z_n_sca_con}
\begin{aligned}
&\left(z_n^\circ\right)^2 + 2z_n^\circ\left(z_n-z_n^\circ\right)-\frac{\left\|\bm{q}_{n+1}^\circ-\bm{q}_n^\circ\right\|^2}{v_0^2} \\
& +\frac{2}{v_0^2}\left[\bm{q}_{n+1}^\circ-\bm{q}_n^\circ\right]^T\left[\bm{q}_{n+1}-\bm{q}_n\right] \ge \frac{\left(T/N\right)^4}{z_n^2} , \quad \forall n\in\mathcal{N}.
\end{aligned}
\end{equation}
Based on the treatment above, we now have the convex version of the problem in~(\ref{eq:problem_4_eqiv}) as
\begin{IEEEeqnarray}{CL} \label{eq:problem_4_eqiv_sca}
	\IEEEyesnumber \IEEEnosubnumber*
	\min_{\substack{ \tilde E_4, \left\{ {\bm{q}_n}, z_n \right\}_{\forall n \in \mathcal{N}} \\ \left\{ {p_{n,k}} \right\}_{ \forall n \in \mathcal{N}, \forall k \in \mathcal{K} } }} & E_1 + \eta_2 \tilde E_4 \\
	{\rm{s.t.}} & (\ref{eq:q_sf}), (\ref{eq:speed}), (\ref{eq:l_nk}), (\ref{eq:pwr_con}), \nonumber \\
	& (\ref{eq:prop_con_4}), (\ref{eq:z_n_sca_con}), \nonumber
	\IEEEeqnarraynumspace
\end{IEEEeqnarray}
as approximated at $ \left\{ \bm{q}_n^\circ, z_n^\circ\right\}_{n\in\mathcal{N}} $. Then, we iteratively solve the problem in the form of~(\ref{eq:problem_4_eqiv_sca}) with updated approximation points through the previously obtained optimum, and thus approach the solution to the original problem in~(\ref{eq:problem_4}).

In the preceding discussions, we have analyzed the decomposed subproblems and proposed the schemes to solve them effectively. When solving each subproblem, the proposed approach is based on the inherent assumption that the variables concerned in other subproblems remain constant. Here, the connection and dependency between the subproblems are unfolded in the following aspects. The offloading decision influences the UAV computation load, which in turn affects UAV-satellite transmission requirements. The time allocation impacts offloading efficiency and UAV-to-satellite transmission feasibility. The beamforming optimization depends on the computed results and available transmission time. Also, the UAV trajectory control affects link quality for both offloading and UAV-satellite transmission, influencing the feasibility of all other subproblems. Accordingly, we can employ the block coordinate descent framework to comprehensively design the algorithm while incorporating all the subproblems. The overall algorithm is summarized in Alg.~\ref{alg}, where $ \varepsilon_0 $ claims the convergence of the algorithm, and $ \varepsilon_{0,2} $, $ \varepsilon_{0,3} $, and $ \varepsilon_{0,4} $ are the thresholds for the convergence of the successive convex approximation procedures when solving the subproblems. Particularly, at the initialization stage, optimization variables can be randomly determined while satisfying the constraints. While for the trajectory, it can be initialized by visiting all the ground nodes by solving it as a traveling salesman problem, with constant flying speed during the time horizon. Then, the subproblems are solved separately so that each group of optimal variables are obtained and updated. The convergence is achieved when the energy decreasing becomes sufficiently small.

As the proposed algorithm follows a block coordinate descent framework, where each subproblem is solved iteratively with fixed settings of others. Then, the convergence is justified in the following aspects. Since each subproblem intends to minimize the energy consumption, the overall energy is monotonically decreased during the iterations. Also, the successive convex approximation adopted within the subproblems is also guaranteed to converge to a local optimum. Therefore, the convergence of the proposed overall block coordinate descent-based energy minimization framework can be established.

Regarding the implementation of the proposed algorithm, the controller can be deployed at the UAV or satellite, collecting the channel information and conducting the algorithm, and feeds back the optimized beamforming, scheduling, and trajectory to the corresponding roles in the system. The channel information is obtained locally at each node and exchanged as communication overhead to assist the algorithm implementation.
Furthermore, the proposed NTN-based integrated communication and computation framework can be conveniently implemented in real-world systems. The framework is compatible with emerging low-earth orbit satellite constellations, where UAVs serve as mobile edge computing nodes. Also, the approach can be deployed in the areas where terrestrial networks are unavailable. The UAV conducts remote computing with on-demand task scheduling and trajectory optimization. As a further note, our algorithm is relatively computationally efficient and can be executed on UAV or satellite onboard processors, or it can be deployed as a cloud-based service on platforms for real-time processing.

\begin{algorithm}[t] \label{alg} %\small
  \caption{Block coordinate descent for energy-efficient integrated communications and computation}
  Randomly initialization of the optimization variables with all constraints satisfied, denoted by $ \left\{ {{\bm{q}_n^0},{\rho_n^0},{\bm{W}_n^0}} \right\}_{\forall n}$, $ \left\{ {{p_{n,k}^0},{\chi _{n,k}^0},{l_{n,k}^0}} \right\}_{ \forall n , \forall k } $\;
  % Initialization: randomly determine the optimization variables on condition that all the constraints are satisfied, while the initial trajectory is determined by solving the travelling salesman problem, $ \left\{ {{\bm{q}_n},{\rho_n},{\bm{W}_n}} \right\}_{\forall n \in \mathcal{N}}, \left\{ {{p_{n,k}},{\chi _{n,k}},{l_{n,k}}} \right\}_{ \forall n \in \mathcal{N}, \forall k \in \mathcal{K} } $ \;
  Derive the deterministic-form reformulated problem in~(\ref{eq:problem_equiv}) by using the Bernstein-type inequalities\;
  \Repeat{$ \resizebox{1.0\hsize}{!}{$E_0 - E\left( \left\{ {{\bm{q}_n^0},{\rho_n^0},{\bm{W}_n^0}} \right\}_{\forall n}, \left\{ {{p_{n,k}^0},{\chi _{n,k}^0},{l_{n,k}^0}} \right\}_{ \forall n , \forall k } \right) < \varepsilon_0$} $}
  {
    $ \resizebox{1.0\hsize}{!}{$E_0 \gets E\left( \left\{ {{\bm{q}_n^0},{\rho_n^0},{\bm{W}_n^0}} \right\}_{\forall n}, \left\{ {{p_{n,k}^0},{\chi _{n,k}^0},{l_{n,k}^0}} \right\}_{ \forall n , \forall k } \right)$} $;
    Solve problem in~(\ref{eq:problem_1}), obtain $ \left\{ {{\chi _{n,k}^{\star}},{l_{n,k}^{\star}}} \right\}_{ \forall n , \forall k } $\;
    % and assign $ \left\{ {{\chi _{n,k}^{0}},{l_{n,k}^{0}}} \right\}_{ \forall n , \forall k } $, update the total energy in~(\ref{eq:E}) as $ E_{0,1} $
    Initialize $ \left\{ \rho_{n}^{\circ} \right\}_{\forall n} $ with $ \left\{ \rho_{n}^{0} \right\}_{\forall n} $\;
    \Repeat{$\resizebox{1.0\hsize}{!}{$ \left| E_{0,2} - E\left( \left\{ {{\bm{q}_n^0},{\rho_n^{\star}},{\bm{W}_n^0}} \right\}_{\forall n}, \left\{ {{p_{n,k}^0},{\chi _{n,k}^{\star}},{l_{n,k}^{\star}}} \right\}_{ \forall n , \forall k } \right)  \right| < \varepsilon_{0,2} $}$}
    {
      $ \resizebox{1.0\hsize}{!}{$E_{0,2} \gets E\left( \left\{ {{\bm{q}_n^0},{\rho_n^0},{\bm{W}_n^0}} \right\}_{\forall n}, \left\{ {{p_{n,k}^0},{\chi _{n,k}^{\star}},{l_{n,k}^{\star}}} \right\}_{ \forall n , \forall k } \right)$} $;
      Formulate the problem in~(\ref{eq:problem_2_sca}) at the point $ \left\{{\rho_n^{\circ}}\right\}_{\forall n} $, and find the solution as $ \left\{\rho_{n}^{\star}\right\}_{\forall n} $\;
      $ \rho_{n}^{\circ} \gets \rho_{n}^{\star} $, $ \forall n\in\mathcal{N} $\;
    }
    Initialize $ \left\{ \bm{W}_{n}^{\circ} \right\}_{\forall n} $ with $ \left\{ \bm{W}_{n}^{0} \right\}_{\forall n} $\;
    \Repeat{$\resizebox{1.0\hsize}{!}{$ \left| E_{0,3} - E\left( \left\{ {{\bm{q}_n^0},{\rho_n^{\star}},{\bm{W}_n^{\star}}} \right\}_{\forall n}, \left\{ {{p_{n,k}^0},{\chi _{n,k}^{\star}},{l_{n,k}^{\star}}} \right\}_{ \forall n , \forall k } \right)  \right| < \varepsilon_{0,3} $}$}
    {
      $ \resizebox{1.0\hsize}{!}{$E_{0,3} \gets E\left( \left\{ {{\bm{q}_n^0},{\rho_n^{\star}},{\bm{W}_n^0}} \right\}_{\forall n}, \left\{ {{p_{n,k}^0},{\chi _{n,k}^{\star}},{l_{n,k}^{\star}}} \right\}_{ \forall n , \forall k } \right)$} $;
      Formulate the problem in~(\ref{eq:problem_3_sca}) at the point $ \left\{{\bm{W}_n^{\circ}}\right\}_{\forall n} $, and find the solution as $ \left\{\bm{W}_{n}^{\star}\right\}_{\forall n} $\;
      $ \bm{W}_{n}^{\circ} \gets \bm{W}_{n}^{\star} $, $ \forall n\in\mathcal{N} $\;
    }
    Initialize $ \left\{ \bm{q}_{n}^{\circ} \right\}_{\forall n} $ with $ \left\{ \bm{q}_{n}^{0} \right\}_{\forall n} $, and $ \left\{ z_{n}^{\circ} \right\}_{\forall n} \ge 0 $\;
    \Repeat{$\resizebox{1.0\hsize}{!}{$ \left| E_{0,4} - E\left( \left\{ {{\bm{q}_n^{\star}},{\rho_n^{\star}},{\bm{W}_n^{\star}}} \right\}_{\forall n}, \left\{ {{p_{n,k}^{\star}},{\chi _{n,k}^{\star}},{l_{n,k}^{\star}}} \right\}_{ \forall n , \forall k } \right)  \right| < \varepsilon_{0,4} $}$}
    {
      $ \resizebox{1.0\hsize}{!}{$E_{0,4} \gets E\left( \left\{ {{\bm{q}_n^0},{\rho_n^{\star}},{\bm{W}_n^{\star}}} \right\}_{\forall n}, \left\{ {{p_{n,k}^0},{\chi _{n,k}^{\star}},{l_{n,k}^{\star}}} \right\}_{ \forall n , \forall k } \right)$} $;
      Formulate the problem in~(\ref{eq:problem_4_eqiv_sca}) at the point $ \left\{{\bm{q}_n^{\circ}}, z_n^{\circ}\right\}_{\forall n} $, and find the solution as $ \left\{\bm{q}_{n}^{\star}, z_n^{\star}\right\}_{\forall n} $, and $ \left\{ p_{n,k}^{\star} \right\}_{\forall n, \forall k} $ \;
      $ \bm{q}_{n}^{\circ} \gets \bm{q}_{n}^{\star} $, and $ z_{n}^{\circ} \gets z_{n}^{\star} $, $ \forall n\in\mathcal{N} $\;
    }
    $ \bm{q}_{n}^{0} \gets \bm{q}_{n}^{\star} $, $ \bm{W}_{n}^{0} \gets \bm{W}_{n}^{\star} $, and $ \rho_{n}^{0} \gets \rho_{n}^{\star} $, $ \forall n\in\mathcal{N} $\;
    $ p_{n,k}^{0} \gets p_{n,k}^{\star} $, $ \chi_{n,k}^{0} \gets \chi_{n,k}^{\star} $, and $ l_{n,k}^{0} \gets l_{n,k}^{\star} $, $ \forall n\in\mathcal{N} $, $ \forall k\in\mathcal{K} $\;
  }
\end{algorithm}

\section{Simulation Results} \label{sec6}

In this section, we present the simulation results to demonstrate the performance of our proposed approach. We consider an area of 1,000~m $\times$ 1,000~m in which 10 ground nodes are randomly placed. The flight altitude of the UAV is fixed at 100~m, and it flies from the initial point (0, 500~m) to the end point (500~m, 0), with the maximum flight speed of 30~m/s throughout the flight. The flight duration is 110~s and is equally divided into 110~time slots. The offloading data amount from the ground nodes is identically set to be 170~Mbits. The maximum computation frequency of the UAV is 6~GHz, with an effective capacitance of 10\textsuperscript{-28}. The produced computation results is proportional to the input data amount with a coefficient of 0.1. The UAV-satellite distance is set 600~km, which remains constant during the time of interest. For the UAV-satellite link, the rain attenuation power loss in decibel has an expectation of $-$8.6~dB and a standard derivation of 0.3~dB. The antenna gains are exploited with their corresponding maximums. The phase in satellite link has an uncertainty with a standard derivation of ${\frac{\text{50}}{\text{360}}}^{\circ}$, and the associated outage probability is 0.2. The satellite is of 8 antennas and the antenna separation is of half-wave length. The simulation parameters are comprehensively listed in Table~\ref{tab}, which are used as the default settings unless otherwise noted. Also, since the performance depends on specified realization of ground node locations, we stick to the same topology, as shown in the following figures, for all the simulation trials and results.

\begin{table}[!htbp] %\small
\caption{Simulation parameters}
\centering
\begin{tabular}{ cc|cc }
\toprule
Notation    & Value & Notation & Value \\
\midrule
% Environmental parameters
$T$                &  $110\,\text{s}$         & $f^\text{max}$                &  $6\,\text{GHz}$ \\
$N$                &  $110$                   & $o$      &  $0.1$ \\
$K$                &  $10$                    & $p^{\text{(UAV)},\max}$ &  $10\,\text{W}$ \\
% Transmission and Frequency parameters
$q_\text{S}$       &  $(0, 500\,\text{m})$    & $B_2$             &  $30\,\text{MHz}$ \\ 
$q_\text{F}$       &  $(500\,\text{m}, 0)$    & $P_0$              &  $79.86$ \\

% Signal and noise levels
$H$                &  $100\,\text{m}$        & $U_\text{tip}$     &  $120$ \\
$V^\text{max}$     &  $30\,\text{m/s}$       & $P_i$             &  $88.63$ \\
$d^{\text{(SU)}}_n$&  $6\times10^5\,\text{m}$  &  $v_0$             &  $4.03$ \\

% Device capabilities
$\beta _0$    &  $-60\,\text{dB}$                   & $s$               &  $0.05$ \\
${G_{t}^{\max}}{G_{r}^{\max}}$               &  $18\,\text{dB}$                    & $d_0$           &  $0.6$ \\
$M$              &  $8$                   & $\rho$            &  $1.225\,\text{kg/m}^3$  \\
$\kappa$          &  $1.38\times10^{-23}\,\text{J/K}$                 & $A$                &  $0.503$ \\
$B_1$             &  $3\,\text{MHz}$       & $\eta_1$     &  $0.1$  \\
$t$ &  $207\,\text{K}$     & $\eta_2$     &  $0.001$  \\

% Energy and motion
$D_k$              &  $170\,\text{Mbits}$         & $E^{\text{(prop)}}$   &  $2\times10^4\,\text{J}$ \\
$\sigma_1^2$       &  $-105\,\text{dBm}$                & $o$               &  $0.1$ \\
$p_k^\text{max}$    &  $1\,\text{W}$      & $\phi^{\text{(SU)}}$               &  $1$\\
$c_0$             &  $100$                   & $c$               &  $3\times10^8\,\text{m/s}$ \\
$f^{\text{(carr)}}$                &  $20\,\text{GHz}$                    & $\gamma$           &  $1\times10^{-27}$ \\
\bottomrule
\end{tabular}\label{tab}
\end{table}

\begin{figure*}[t] 
  \centerline{
  \subfigure[UAV trajectory with speed variation.]{
    \label{fig:T_speed} %% label for first subfigure
    \includegraphics[width=6.0cm]{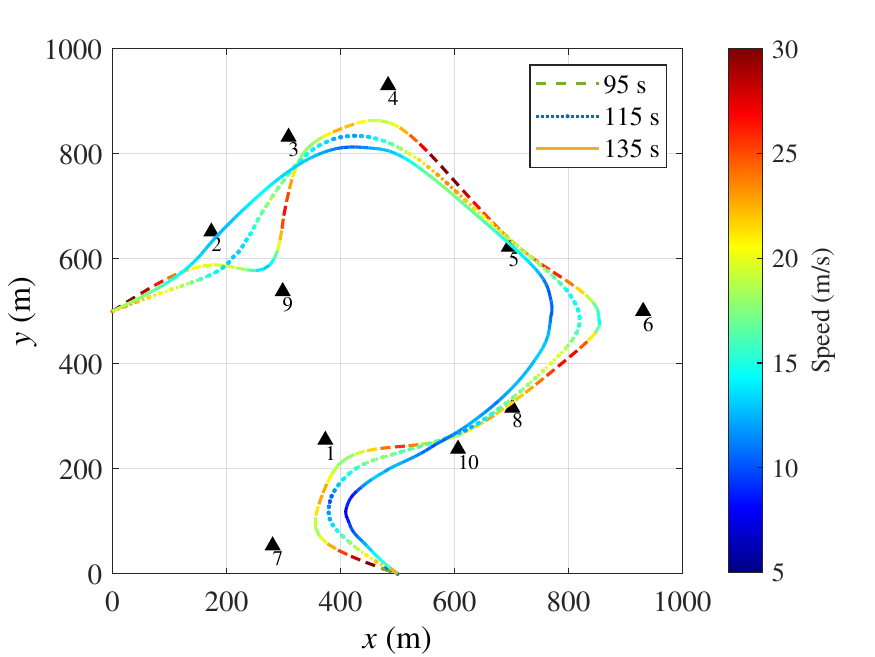}}
  \subfigure[Offloading data amount during UAV flight.]{
    \label{fig:T_off} %% label for second subfigure
    \includegraphics[width=6.0cm]{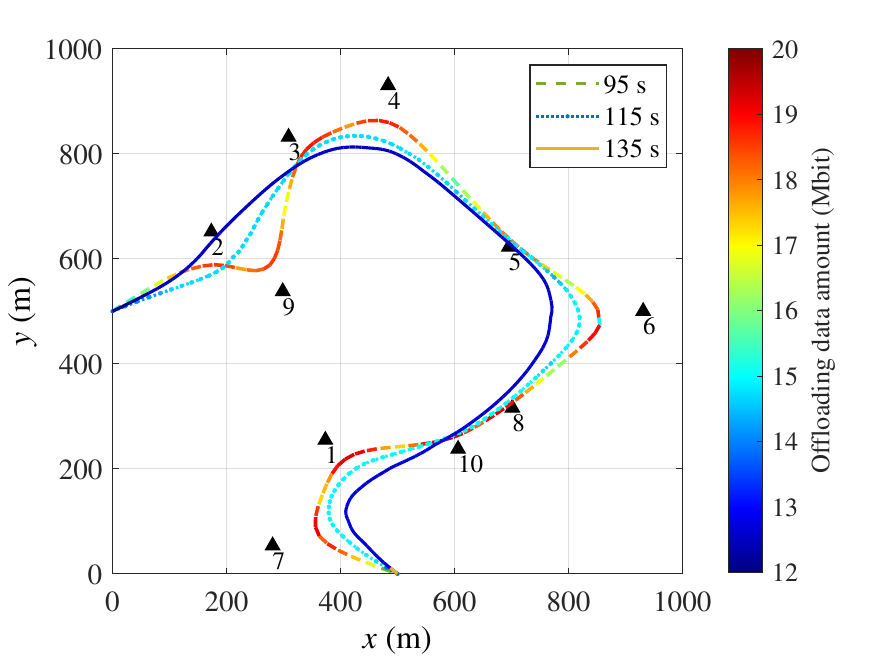}}
    \subfigure[Energy consumption breakdown.]{
    \label{fig:T_energy} %% label for first subfigure
    \includegraphics[width=6.0cm]{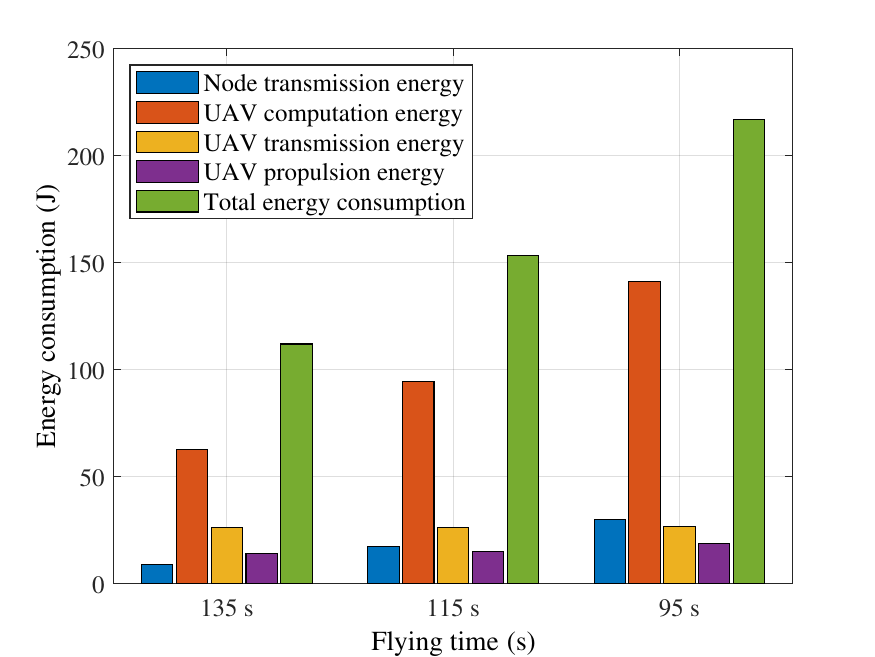}}
  }
  \caption{Performance under different time lengths.}
  \label{fig:T_perfm} %% label for entire figure
\end{figure*}

In Fig.~\ref{fig:T_perfm}, we show the performance in terms of trajectory, offloading, and energy with different time settings. Particularly, in Fig.~\ref{fig:T_speed}, we can see the average flying speed becomes lower with longer time window, and vice versa in the case of shorter allowed time. This is as expected given the natural relationship between flying mission and time constraint. An interesting observation is that, with longer time given, the UAV flies in a more ``relaxed'' manner such that the overall flying trajectory is smoother and the change in speed is less evident. While in contrast, when the allowed time is short, the UAV flying approaches closer to the ground nodes, which incurs longer flying distance and thus the change in flying speed also becomes more dramatic. Since the most energy-efficient flying speed under given parameters regarding propulsion energy is about 12~m/s, we can see that, correspondingly, when more time is allowed, the flying speed approaches such an optimal speed so as to save the propulsion energy. Accordingly, in Fig.~\ref{fig:T_off} where the offloading data amount during the flight is specified, we can see that the change in offloading amount is less evident with longer allowed time. In contrast, when the time horizon is shortened, the offloading data amount changes significantly while flying. This is partially because the shorter time window for offloading induces more uneven mission distribution along time. Another contributing factor is that the induced trajectory more closely approaches the ground nodes, and thus the nodes tend to offload more when the UAV has a shorter distance with them to save the transmission energy of their own. Fig.~\ref{fig:T_energy} shows the overall energy consumption along with the weighted components. As expected, with shorter time, the ground nodes need to offload with higher transmission rate and the UAV needs to fly faster and conduct edge computing with higher frequency, and thus the energy due to ground nodes transmission, UAV flying and computation, and the overall energy increase significantly. Particularly, the UAV computation energy manifests the most striking increase due to the cubic relationship between the computation energy and the computing frequency. Meanwhile, changes in energy for UAV-satellite transmission are less noticeable with changes in time length, as the significant distance between the UAV and satellite renders changes in the ground environment negligible. Overall, the results in Fig.~\ref{fig:T_perfm} indicate that, with a shorter time to finish the offloading task, the UAV needs to more closely approach the ground nodes and induces more dynamic changes in flying speeds and offloading amounts while flying, and thus the total energy consumption is significantly increased.

\begin{figure*}[t] 
  \centerline{
  \subfigure[UAV trajectory with speed variation.]{
    \label{fig:FD_speed} %% label for first subfigure
    \includegraphics[width=6.0cm]{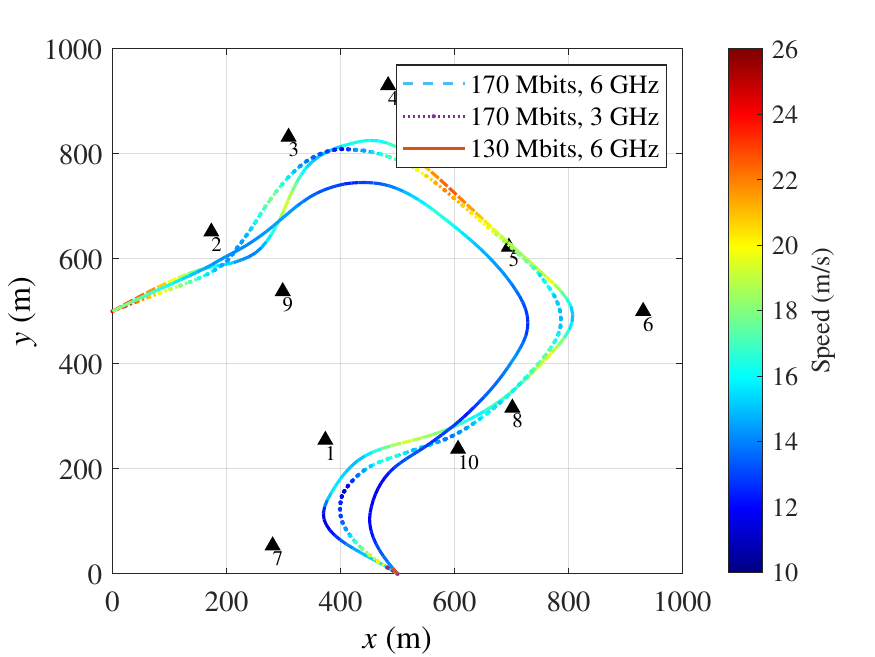}}
  \subfigure[Offloading data amount during UAV flight.]{
    \label{fig:FD_off} %% label for second subfigure
    \includegraphics[width=6.0cm]{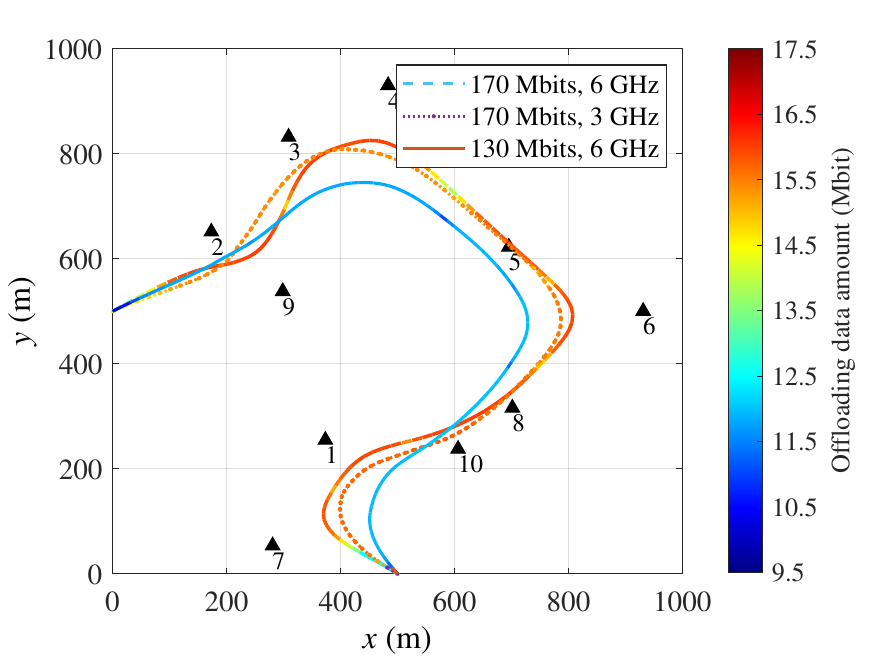}}
    \subfigure[Energy consumption breakdown.]{
    \label{fig:FD_energy} %% label for first subfigure
    \includegraphics[width=6.0cm]{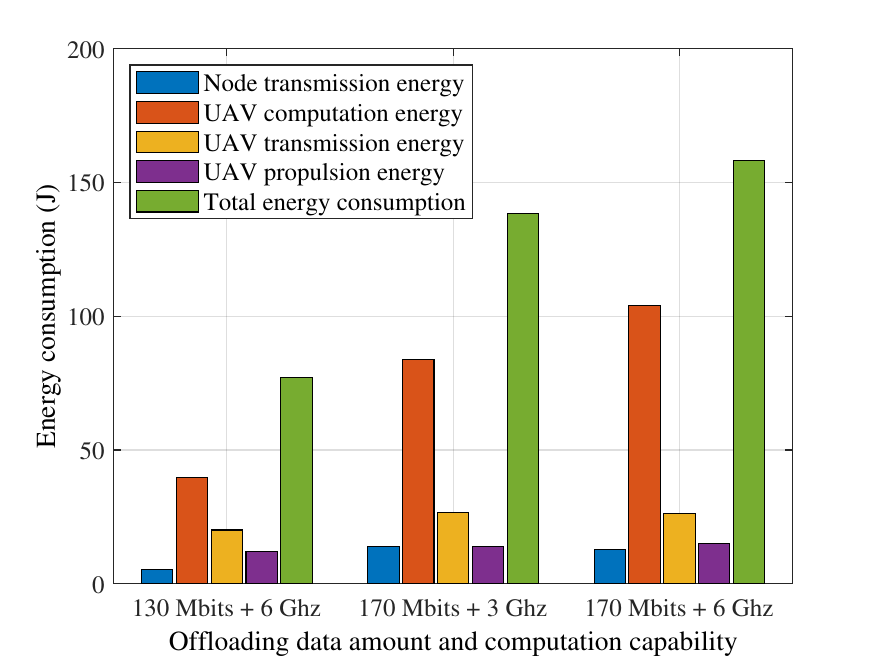}}
  }
  \caption{Performance under different data amounts and computation capabilities.}
  \label{fig:FD_perfm} %% label for entire figure
\end{figure*}

In Fig.~\ref{fig:FD_perfm}, we demonstrate the performance with different offloading data amounts and computation capabilities. Specifically, from Fig.~\ref{fig:FD_speed} showing the trajectories under different settings, we can see that with an increasing data offloading requirement, the flying trajectory approaches the ground nodes more closely, facilitating the offloading operations. Additionally, the increased offloading data amount results in a longer flying distance, which in turn necessitates higher flying speeds during most of the time intervals. In the meantime, as indicated in Fig.~\ref{fig:FD_off}, with larger overall data amounts, the more aggressive flying is associated with increased offloading during the flight. As a result, the energy consumed in different aspects as well as the overall energy consumption are evidently increased, as shown in Fig.~\ref{fig:FD_energy}, where the UAV computation energy is significantly increased. Moreover, increasing the maximum computation frequency results in a more dynamic flight trajectory, bringing the UAV closer to the ground nodes, as shown in Fig.~\ref{fig:FD_speed}. Consequently, offloading becomes less balanced under higher computation capabilities, as seen in Fig.~\ref{fig:FD_off}, with locations closer to the ground nodes experiencing higher data amounts. Further, since the UAV needs to conduct higher-frequency computation for the locations with higher offloading amounts, the energy for computation can be increased substantially given the cubic increase of energy with respect to computation frequency. Simultaneously, as the UAV approaches the ground nodes more closely, the energy required for offloading transmission decreases, conserving energy for the ground nodes. Therefore, the overall energy consumption is increased with higher allowed computation frequency.

\begin{figure*}[t]  %\vspace{-2pt}
  \centerline{
  \subfigure[UAV trajectory with speed variation.]{
    \label{fig:Comp_speed} %% label for first subfigure
    \includegraphics[width=6.0cm]{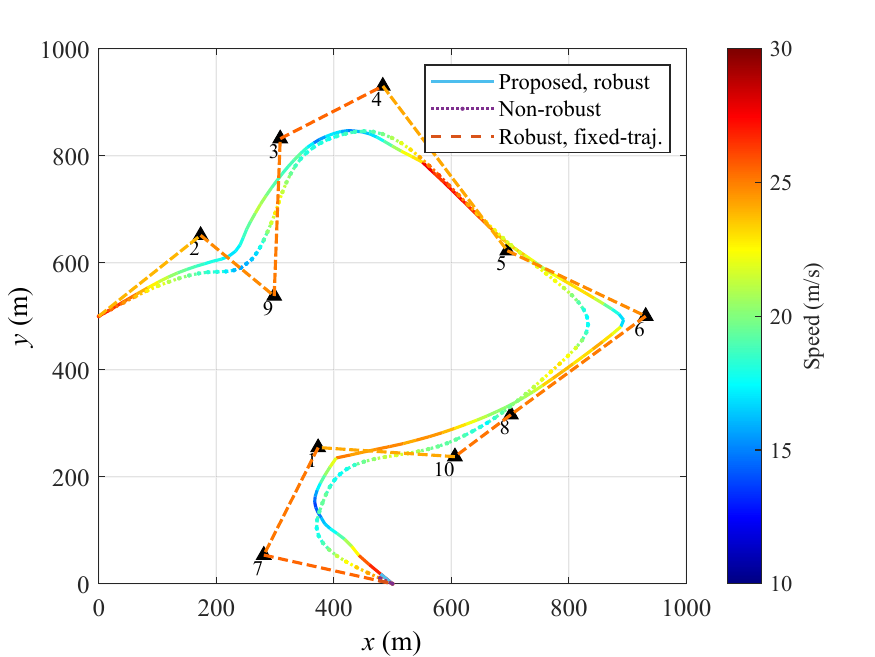}}
  \subfigure[Offloading data amount during UAV flight.]{
    \label{fig:Comp_off} %% label for second subfigure
    \includegraphics[width=6.0cm]{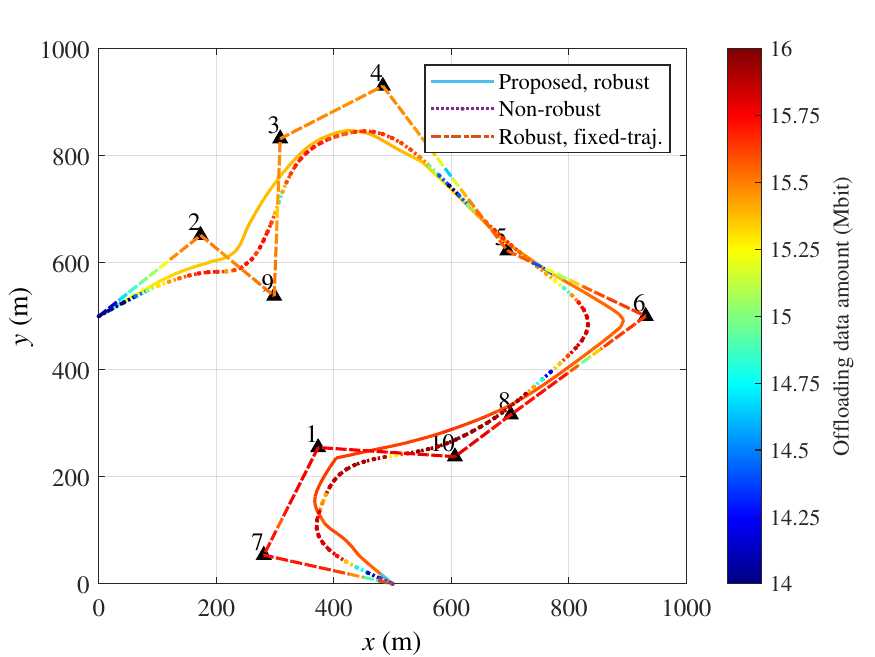}}
    \subfigure[Energy consumption breakdown with respect to computation data amount.]{
    \label{fig:Comp_energy} %% label for first subfigure
    \includegraphics[width=6.0cm]{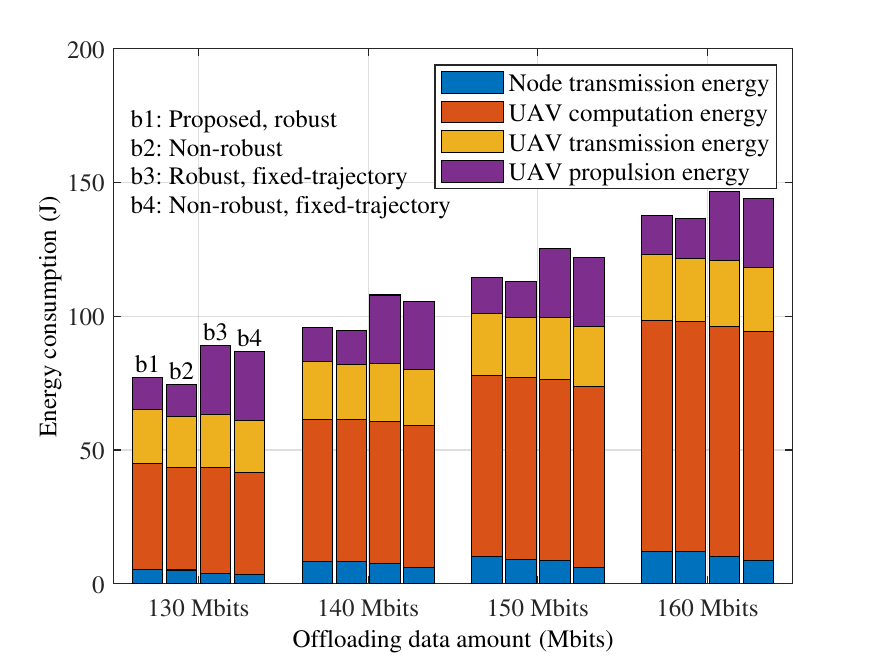}}
  }
    \centerline{
  \subfigure[Energy consumption in different aspects while flying.]{
    \label{fig:Comp_component} %% label for first subfigure
    \includegraphics[width=8.5cm]{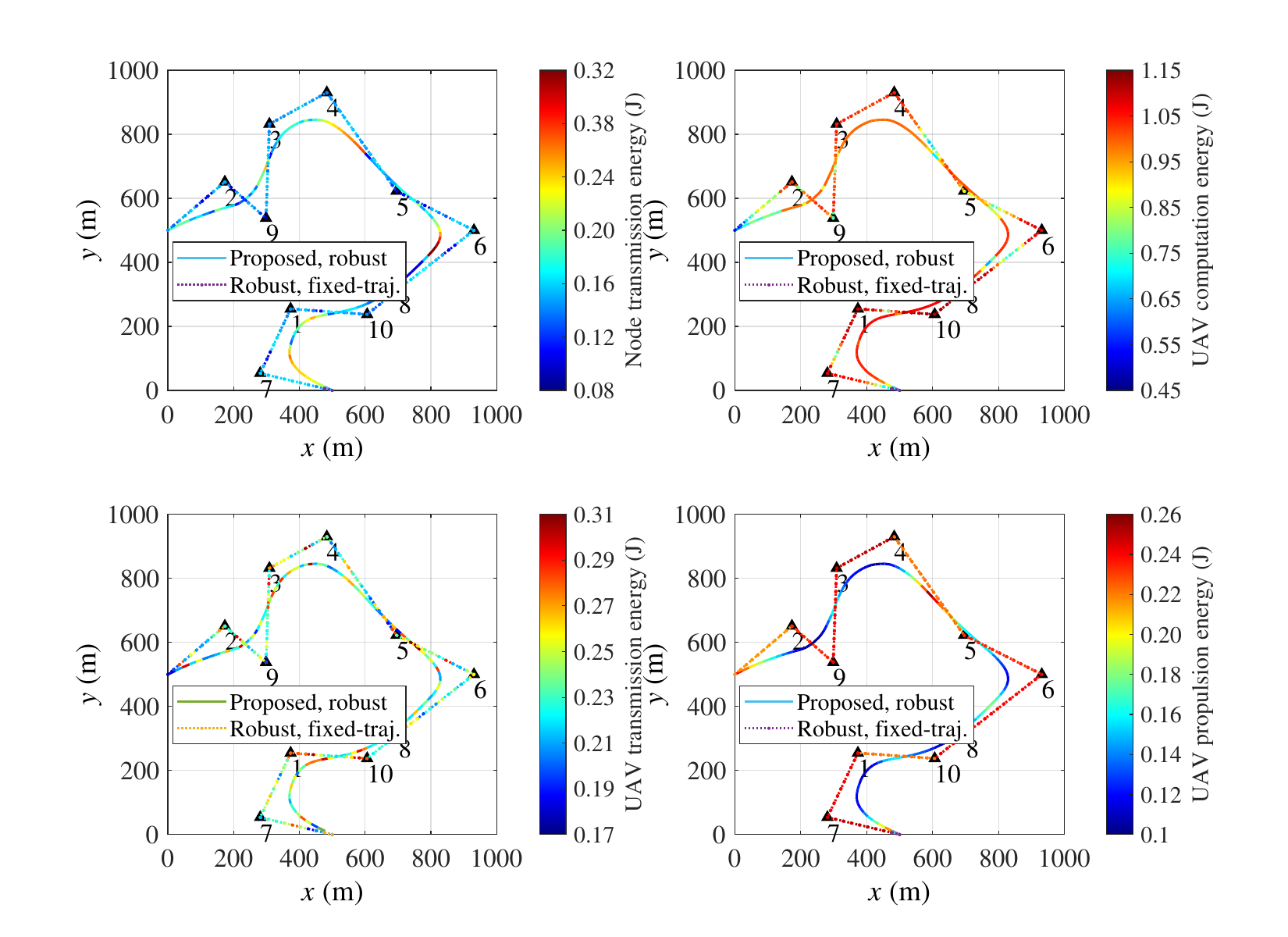}}
  \subfigure[Histogram of results forwarding completion ratio with different outage probabilities.]{
    \label{fig:Comp_distribution} %% label for second subfigure
    \includegraphics[width=8.5cm]{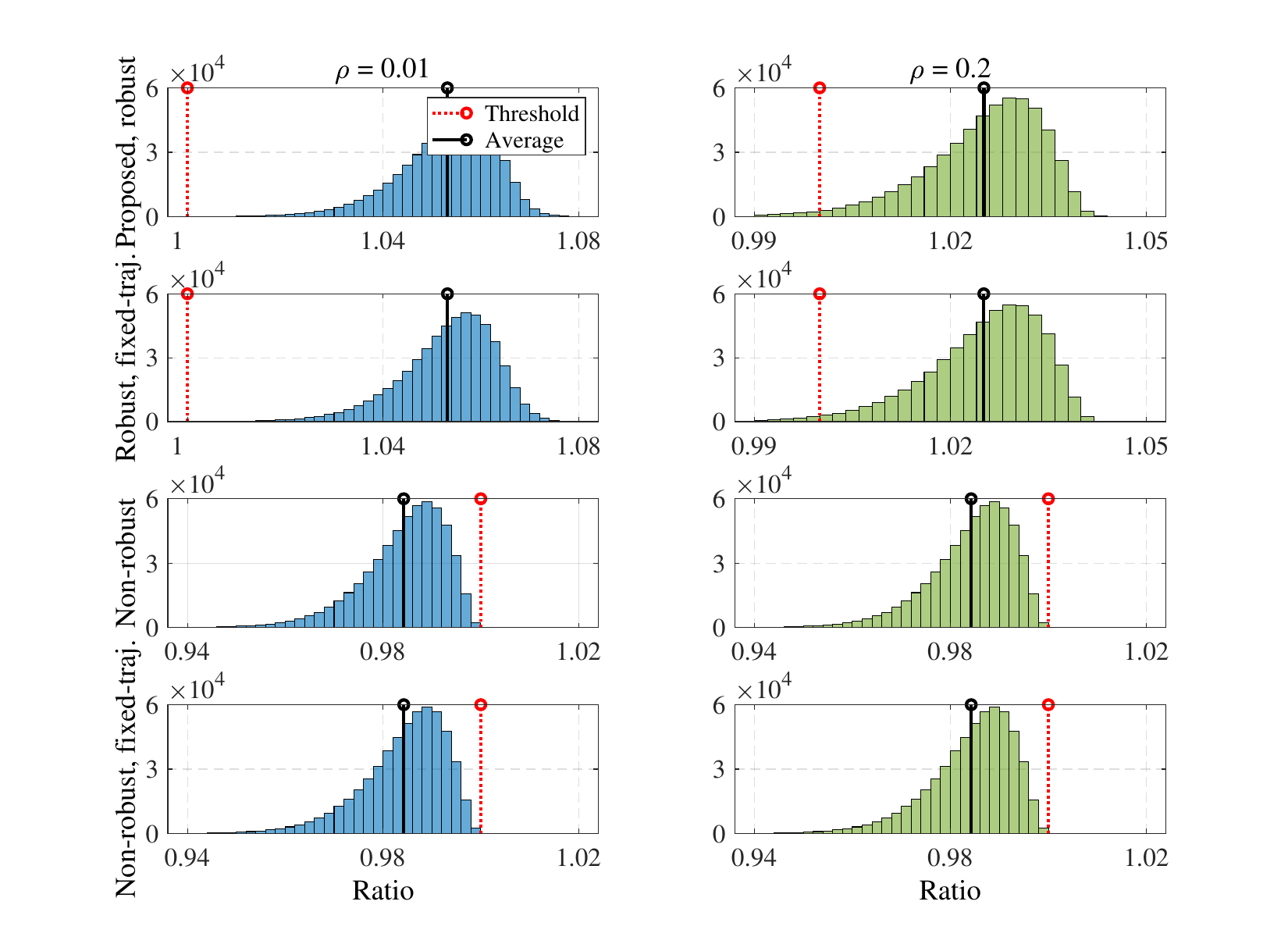}}
  }
  \caption{Performance comparison under different schemes.}
  \label{fig:Comp_perfm} %% label for entire figure
\end{figure*}

In Fig.~\ref{fig:Comp_perfm}, we show the performance under different proposals, where the non-robust scheme solves the problem using the constraints in~(\ref{eq:result_con}) rather than the probabilistic counterparts in~(\ref{eq:chance_con}), regardless of the angular uncertainties. Also, the schemes on condition of fixed trajectory are the counterparts of the robust and non-robust schemes that simply connect the ground node locations through traveling salesman problem with optimized flying speeds. In Figs.~\ref{fig:Comp_speed} and~\ref{fig:Comp_off}, the flying trajectories are shown along with the flying speeds and offloading amounts. Since the robust and non-robust schemes under a fixed trajectory share the same path, we only show cases using the robust design. The results indicate that the proposed robust scheme induces a trajectory that tends to be closer to the ground nodes as compared with the non-robust scheme. Accordingly, the flying distance becomes slightly longer and thus the change in speed is more dynamic under the robust scheme. For the fixed-trajectory case, the flying distance is much longer without optimized flying path, leading to faster flying speed. Meanwhile, for the offloading operations, the proposed robust scheme generally has more stable offloading operation during the flight, and in contrast, the offloading under the non-robust scheme changes more dynamically. This is due to the fact that the robust scheme needs to jointly consider the offloading as well as the outage-constrained results forwarding transmissions, thus the offloading is conducted in a smoother manner. For the non-robust scheme, the offloading and result forwarding are conducted by adapting to the location and environment, and thus the changes in offloading amount are more evident. In the fixed-trajectory case, the increased flying distance provides more space to distribute offloading amounts during the flight, thus alleviating changes in offloading data amounts

In Fig.~\ref{fig:Comp_energy}, we compare the energy consumption along with its components with respect to offloading data amount under different schemes. As expected, with higher offloading data amount, energy consumption increases in every aspect and overall for all schemes. Particularly, the robust schemes induce some slightly higher energy consumption as compared with their non-robust counterparts, as the outage probability under the robust scheme induces different transmission behavior at the UAV-satellite link and flying trajectories. For the cases with fixed trajectory, they generally induce higher energy consumption as compared with the counterpart schemes with optimized trajectory. This is mainly due to the higher propulsion energy resulting from the former cases, where both the flying distance and speed are higher. Moreover, in Fig.~\ref{fig:Comp_component}, we present a detailed view of the energy consumption for different components during flight, focusing on robust designs with various trajectories for clearer demonstration. Particularly, for the ground node transmission energy, we can see that since the fixed-trajectory flying approaches the ground nodes, their energy can be significantly saved. Additionally, because the robust trajectory keeps the UAV relatively farther from the nodes, changes in energy consumption during the flight are more pronounced. For the UAV computation energy, the more conservative trajectory under robust scheme leads to intensive computation, resulting in less evident energy fluctuations. For the UAV transmission to forward the computation results to the satellite, since the two schemes both consider robustness and the total size of the results are the same, the energy consumption in this respect is similar across both schemes. For the propulsion energy, the two schemes, as expected, manifest distinct performance, as the longer flying distance and higher flying speed under the fixed-trajectory scheme induce significantly higher propulsion energy. In a general sense, we can see the trade-off in terms of the UAV propulsion energy and node transmission energy with different trajectory designs. The higher energy expenditure on UAV propulsion with a fixed trajectory helps conserve energy for the ground nodes. Therefore, we can achieve different designs to fit different network settings and service requirements.

In Fig.~\ref{fig:Comp_distribution}, we evaluate the performance regarding the chance-constraint guarantee in the presence of uncertainties under different schemes. Here, we calculate the ratio between the actual transmitted data amount and the threshold amount. We then provide a histogram of these ratios, obtained from UAV-satellite channels randomly generated according to the predefined distribution. Then, the threshold equals one and determines whether the result forwarding transmission is successfully completed. The results show that the robust schemes ensure the outage probability is met, with an average completion ratio consistently above one. Specifically, when the outage probability is higher, allowing some more relaxed guarantee regarding the completion, the average completion ratio is also lowered, but the overall performance still strictly guarantees the outage probability. For the non-robust schemes, we can see that they fail the outage probability, as nearly all the trials have a lower-than-one completion ratio given the randomness in channel conditions. Recall the results in Fig.~\ref{fig:Comp_energy}, which indicate that the robust designs have slightly higher energy consumption. This additional energy expenditure is the price paid to combat uncertainties, ensuring guaranteed performance in uncertain environments.

\section{Conclusion} \label{sec7}

In this paper, we investigate the NTN-based integrated communication and computation, considering network uncertainties, and aim to minimize the energy consumption in terms of data offloading, computation, results forwarding, and UAV propulsion with a robustness guarantee. Results indicate that the proposed solution adapts to the varying requirements of time length, data amounts, and computation capacity, effectively saving energy and ensuring robust performance despite uncertainties. Moreover, the proposed solution has the scalability to exploit different trajectories to achieve effective tradeoff regarding the energy consumption of different parties in the network.

\bibliographystyle{IEEEtran}
\bibliography{main}

\end{document}